\documentclass[twocolumn]{aa}
\usepackage[utf8]{inputenc}
\usepackage{graphicx}
\usepackage{grffile}
\usepackage{hyperref}

\usepackage{amsmath,amssymb}
\usepackage{color}
\usepackage{ulem}

\newcommand*\df{\mathop{}\!\mathrm{d}}
\newcommand{\n}{\nonumber}
\newcommand*\vr{\mathbf{r}}
\newcommand*\vk{\mathbf{k}}
\newcommand*\LSR{\mathrm{LSR}}
\newcommand{\degree}{\circ}
\newcommand{\HI}{\text{H}\textsc{i}\xspace}
\newcommand{\HII}{\text{H}\textsc{ii}\xspace}
\newcommand{\Htwo}[0]{\ensuremath{\mathrm{H}_2}}%

\makeatletter
\newcommand*{\rom}[1]{\expandafter\@slowromancap\romannumeral #1@}
\makeatother

\defcitealias{bissantz2003}{BEG03}
\defcitealias{sormani2015}{SBM15}
\defcitealias{green2019}{GSZ19}
\defcitealias{leike2022}{LAE22}

\title{Bayesian inference of three-dimensional gas maps}

\subtitle{II. Galactic HI}
\date{}

\author{
P.~Mertsch\inst{\ref{inst1}}
\and
V.H.M.~Phan\inst{\ref{inst2}}
}

\institute{
Institute for Theoretical Physics and Cosmology (TTK), RWTH Aachen University, Sommerfeldstr. 16, 52074 Aachen, Germany, \email{pmertsch@physik.rwth-aachen.de}\label{inst1}
\and
Institute for Theoretical Physics and Cosmology (TTK), RWTH Aachen University, Sommerfeldstr. 16, 52074 Aachen, Germany, \email{vhmphan@physik.rwth-aachen.de}\label{inst2}
}

\abstract{
The 21 cm emission from atomic hydrogen (\HI{}) is one of the most important tracers of the structure and dynamics of the interstellar medium. Thanks to Galactic rotation, the line is Doppler shifted and, assuming a model for the velocity field, data from gas line surveys can be deprojected along the line of sight. However, given our vantage point in the Galaxy, such a reconstruction suffers from a number of ambiguities. Here, we argue that those can be cured by exploiting the spatial coherence of the gas density that is implied by the physical processes shaping it. We have adopted a Bayesian inference framework that allows reconstructing the three-dimensional map of \HI{} and quantifying its uncertainty. We employ data from the HI4PI compilation to produce three-dimensional maps of Galactic \HI{}. The reconstructed density shows structure on a variety of scales. In particular, some spurs and spiral arms can be identified with ease. We discuss the morphology of the surface mass density and the radial and vertical profiles. The reconstructed three-dimensional \HI{} densities are available at \href{https://doi.org/10.5281/zenodo.5956696}{this https URL}.
}


\begin{document}

\maketitle

\section{Introduction}

Hydrogen is the most abundant element in the universe by far. All heavier elements are formed from it, either by primordial or stellar nucleosynthesis. In the Galaxy, most of the hydrogen is locked up in stars, with its gaseous fraction only amounting to about $10 \, \%$ of its stellar mass~\cite{sparke2006}. This interstellar gas, however, forms the reservoir from which new stars are formed. Its distributions, properties, and dynamics are therefore of paramount importance for understanding and modelling star formation. On larger scales, observations of the gas density in the Galaxy have important consequences for questions of galaxy structure and galaxy evolution~\citep{kewley2019}. In high-energy astrophysics, the gas distribution is an important input for the modelling of diffuse emission in gamma-rays which is primarily due to the production of neutral pions by interactions of non-thermal cosmic rays with the interstellar gas~\cite{tibaldo2021}. 

More than half of the gaseous hydrogen in the Galaxy is in the form of atomic hydrogen (\HI{}), which dominates the cold phases of the interstellar medium (ISM). The mass of molecular hydrogen (\Htwo{}) is about half of that and is located mostly in molecular clouds. Ionised hydrogen (\HII{}) is found in the warm and hot phases of the ISM, that is, mostly in the halo. Because of the low temperature of the phases carrying the \HI{}, most of it will be in the atomic ground state, such that there is no emission by transition from excited states. Instead, the hyperfine transition of the hydrogen ground state has become a standard diagnostic tool for a broad range of astrophysical disciplines. Since the electric dipole transition is forbidden, the magnetic dipole transition has a very long lifetime of $\sim 3.5 \times 10^{14} \, \text{s}$, but the low collision rates in the ISM guarantee that the higher state can de-excite radiatively. In addition, the high column densities available on Galactic and extragalactic scales guarantee that observable emission is accumulated. The emission occurs at a frequency of $1.4204058 \, \text{GHz}$ or a wavelength of $\sim 21 \, \text{cm}$, which gives the emission line its name. Because of the long lifetime, the intrinsic line width is very narrow such that the 21 cm line lends itself to velocity spectroscopy.

Ever since the discovery of the 21 cm line in space \citep{muller1951,ewen1951}, surveys of the Galactic \HI{} distribution have been carried out with increasingly higher sensitivity and larger sky coverage. One concrete example of such surveys is the Leiden/Argentine/Bonn Survey (LAB Survey, \citealt{kalberla2005}), which over the last 20 years has been the de facto source of information on \HI{} in the Galaxy. The LAB Survey has revealed the flared, warped, and multi-arm spiral structure of the \HI{} disk \citep{levine2006a,levine2006b}. Recently, a new all-sky survey of the 21 cm line emission that combines different data sets for both low and high Galactic latitude data was published by the HI4PI collaboration \citep{HI4PI2016}. This survey provides the brightness temperature obtained from the 21 cm line emission in a three-dimensional grid of Galactic longitude $\ell$, Galactic latitude $b$, and the radial velocity $v_{\LSR}$ with respect to the local standard of rest. For a given gas flow model, that is, a model of $v_{\LSR}=v_{\LSR}(\vr)$ for the Milky Way, the three-dimensional gas density can in principle be reconstructed from the gas brightness temperature \citep{kulkarni1982,burton1986,nakanishi2003,kalberla2008,marasco2017}.  

There are a few complications that degrade the quality of the deprojection from an $\ell{}bv$-data cube of brightness temperature to the $xyz$-cube of gas density. First, there are two distance solutions inside the solar circle. It is not clear how we can determine the fractions of the emission that are at near and far distances. This effect is known as the kinematic distance ambiguity. In addition, the line of sight close to the Galactic centre (longitude $\ell \simeq 0^\degree$) and anti-centre (longitude $\ell \simeq \pm 180^\degree$) directions exhibit little to no radial velocity and therefore lack kinematical resolution. This means that all emission piles up around $v_{\LSR}=0 \, \text{km} \, \text{s}^{-1}$ and cannot be deprojected along the line of sight. Finally, peculiar velocities, that is, random motions of gas in addition to the large-scale gas flow, for instance due to stellar winds, supernova explosions, or spiral structures, perturb the smooth mapping of distance to radial velocity. These perturbations become visible as artefacts in the deprojected gas maps. In this case, the distribution of gas is smeared out along the line of sight, leading to the well-known finger-of-god effect.

Historically, the pioneering work of \citet{westerhout1957} and \citet{oort1958} provided parts of the face-on Galactic \HI{} density map using the 21 cm line data. Ever since, a number of studies have tried to improve on these early studies, oftentimes circumventing some of these above-mentioned difficulties by limiting the scope of the deprojection. Using data from the LAB survey, \citet{levine2006a,levine2006b} attempted to uncover both the spiral structure and the radial and vertical distributions. For the velocity model, they assumed circular rotation with a constant velocity of $220 \, \text{km} \, \text{s}^{-1}$. They limited themselves to regions outside the solar circle where there is no kinematic distance ambiguity and lines of sight far enough away from $\ell \simeq 0^\degree$ and $\ell \simeq \pm 180^\degree$. However, the effect of peculiar velocities is still very much visible in terms of finger-of-god effects. This limited the authors in their attempt to identify large-scale structures; in order to identify spiral structures, some more intricate analysis was necessary. One way to limit the influence of low-intensity features that are rather spread out in velocity is to focus only on the densest regions. Such a study has recently been performed by \citet{koo2017}, also employing LAB data. They also constrained themselves to regions outside the solar circle and to sight-lines with finite radial velocities in a circular rotation model. For each sight-line, local maxima of the velocity spectrum above a certain threshold were identified and resolved into individual cloudlets. The resulting density of these emission regions was much better resolved than the overall density, but this comes at the price of only reconstructing part of the \HI{} density. The authors estimate that their surface density map contains $\lesssim 10 \%$ of the total \HI{} gas in the Galaxy. 

Attempts to reconstruct most of the Galactic \HI{} density have been presented by \citet{nakanishi2003} and \citet{nakanishi2016}. Again based on LAB data, the authors attempted to break the kinematic distance ambiguity inside the solar circle by employing a variant of the double-Gaussian method (e.g.~\citealt{clemens1988}). If we assume a parametrised vertical profile and the same height of the gas distribution at the near and far distance, the latitudinal profile at a certain distance should be the sum of two contributions, one narrower and one wider in latitude. When this is fit with the parametrised profile, the emission observed at a certain $v_{\LSR}$ can be broken down into contributions at the near and far distance. Emission close to the Galactic centre or anticentre directions still needed to be excluded. While there are some differences between the earlier and later analysis, both reconstructions show a deficit in gas density in the solar circle. In addition, the density is markedly smeared out along the sight lines. 

Two other studies are noteworthy here, even though they do not attempt a direct reconstruction of the three-dimensional gas density from line survey data. Focussing on regions inside the solar circle, \citet{marasco2017} assumed spherical symmetry for the gas distribution, at least separately in the half-planes corresponding to positive and negative longitudes. Identifying the tangent points in velocity diagrams, that is, the maximum velocities with significant emission for a given line of sight, they could identify the emissivity for all radii $r \leq R_{\odot}$ where $R_{\odot}$ is the distance of the Sun from the Galactic centre. While only giving some (azimuthal) average, the radial profiles are still useful for studying some radial dependences. A different approach altogether has been presented by \citet{johannesson2018}, who presented a parametric model of \HI{} (and \Htwo{}) fitted to gas line data. While a large number of model parameters can be determined in this way, the model is not flexible enough to reproduce all of the observed emission. The gas density therefore should be considered a lower bound on the actual gas density in the Galaxy.

All these previous attempts ignore an important physical constraint that the reconstructed gas density needs to satisfy and that can be exploited in reconstructing the gas density. All processes determining the gas distribution lead to spatial correlations, be it on large scales (spiral arms from density waves; e.g.~\citealt{shu2016}, disk profile from gravitational collapse) or small scales (turbulence of the ISM; e.g.~\citealt{kolmogorov1941}). Some theoretical arguments and numerical simulations suggest that the statistics of the three-dimensional gas density is log-normal~\citep{nordlund1999,ostriker2001}. Such a log-normal field can be characterised by the two-point correlation in coordinate space. Under the additional assumption of (spatial) stationarity and isotropy, the two-point function in harmonic space becomes a function of the wave number only, the power spectrum. Reconstructing the gas density under the constraint of a known or to be determined correlation structure is a Bayesian inference problem. The Bayesian nature allows us to reconstruct not only the gas density, but also to obtain an estimate of its uncertainty. This method has been successfully applied to a deprojection of the three-dimensional distribution of \Htwo{} in the Galaxy~\citep{mertsch2021b}.

For the problem at hand, a number of the above-mentioned short comings can be overcome when the correlations are taken into account. For instance, the kinematic distance ambiguity can be resolved by demanding that the reconstructed gas density globally satisfy the correlation structure. These correlations can also help where no distance information is available, that is in the directions of $\ell = 0^{\circ}$ and $\ell = \pm 180^{\circ}$. Given that the gas density model is probabilistic, the uncertainty will naturally be larger in spatial regions for which the data are less constraining. Finally, the data model employed takes the presence of noise into account, thus providing some denoising of the observational data.

In the following, we discuss our method and the results of our application to HI4PI data. In Sect.~\ref{sec:method}, we briefly review the data used, introduce the gas flow models employed, and explain the Bayesian reconstruction framework. Our results are presented in Sect.~\ref{sec:results} and are compared with the results of previous analyses. We provide a short summary and outlook in Sect.~\ref{sec:summary}.

\section{Method}
\label{sec:method}

\subsection{Survey data}

\begin{figure*}[th!]
\centering
\includegraphics[width=7.0in, height=3.2 in]{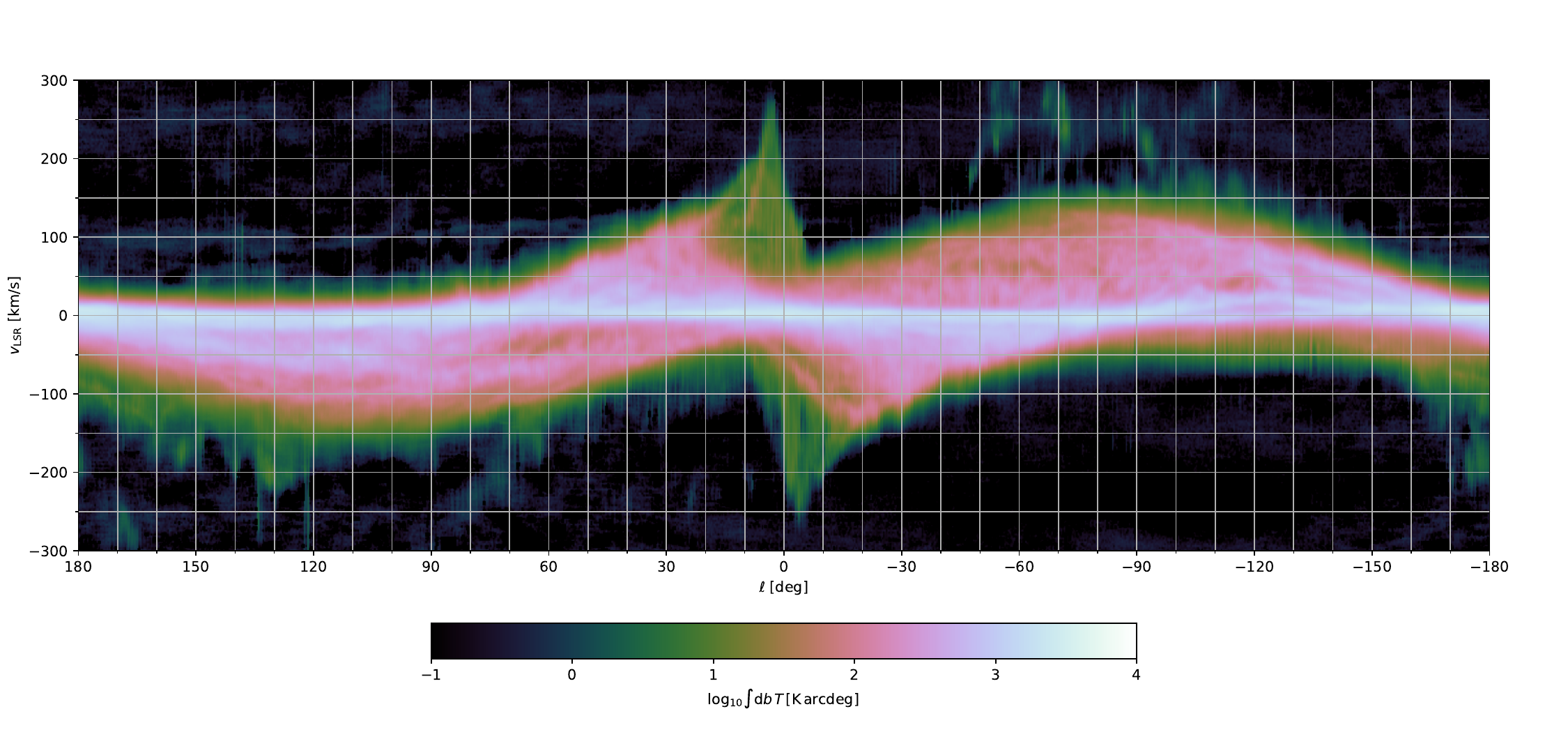}
\caption{$\ell$-$v$-diagram of Galactic 21 cm line emission. The diagram has been obtained by integrating the degraded HEALPix maps with $N_{\rm side}=128$ over latitude (see text for more details).}
\label{fg:l-v-diagram}
\end{figure*}

\begin{figure*}[th!]
\centering
\includegraphics[scale=1]{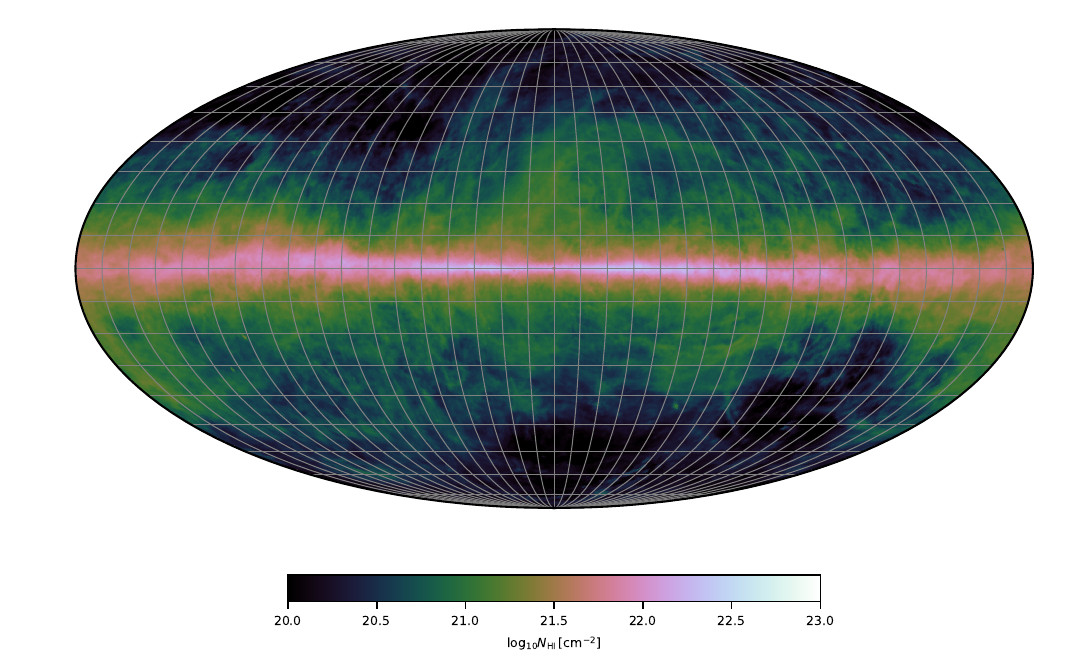}
\caption{All-sky column density of \HI{} obtained from HI4PI data using Eq. \ref{eq:column} for the velocity range $-320\lesssim v_{\LSR} \lesssim 320 \, \text{km} \, \text{s}^{-1}$. The map is presented in Galactic coordinates using a Mollweide projection. The graticule lines are spaced by $10^{\circ}$.}
\label{fg:all-sky-NHI}
\end{figure*}

The reconstruction of the \HI{} density was carried out using HI4PI data \citep{HI4PI2016} that were merged from data of the 21 cm line spectra of two individual surveys, that is, the Effelsberg-Bonn \HI{} Survey (EBHIS; \citealt{kerp2011,winkel2016a}) and the Galactic All-Sky Survey (GASS; \citealt{mccluregriffiths2010,kalberla2010,kalberla2015}). Both surveys have rather similar angular and spectral resolutions (see the survey parameters presented in Tbl.~1 of \citealt{HI4PI2016}). Since EHBIS covers the declination range of $\delta\leq 1^\circ$ and GASS has data for $\delta\geq -5^\circ$, the combined HI4PI survey provides the full-sky data. We downloaded and assembled the prepared data in the form of HEALPix maps with $N_{\rm side}=1024$~\citep{gorski2005} from the Strasbourg astronomical Data Center\footnotemark.

\footnotetext{\url{http://cdsarc.u-strasbg.fr/ftp/J/A+A/594/A116/HPX/}} 

For the spatial resolution we aim for, a coarser representation of the survey data is sufficient. We, therefore, degraded the HEALPix maps from \mbox{$N_{\rm side}=1024$} to $N_{\rm side}=128$ for the reconstruction. More importantly, even though the survey data consists of 933 maps, that correspond to the velocity range from $-600 \, \text{km} \, \text{s}^{-1}$ to $600 \, \text{km} \, \text{s}^{-1}$, we focused only on the velocity range \mbox{$-320 \lesssim v_{\LSR} \lesssim 320 \, \text{km} \, \text{s}^{-1}$} as in the reconstruction of \Htwo{} presented in \citet{mertsch2021b}. We also masked out the combination of the latitude and longitude ranges $-60^{\degree} \leq b < -20^{\degree}$ and $270^{\degree} \leq \ell < 330^{\degree}$ in order to remove contamination from the Magellanic Clouds.

We show as an example the degraded data cube of 21 cm line emission integrated over latitude, also known as the $\ell$-$v$-diagram, in Fig.~\ref{fg:l-v-diagram}. The all-sky map of column density is shown in Fig.~\ref{fg:all-sky-NHI}.

\subsection{Gas flow model}
\label{sec:velocity}

The distribution of \HI{} brightness temperature in $\ell$, $b$, and $v_{\LSR}$ depends on both the three-dimensional density of the gas and its velocity along the line of sight. This means that the reconstruction of the gas density requires knowledge about the gas motion as briefly mentioned above. Even though it is commonly assumed that the gas follows purely circular motions around the Galactic centre, this assumption is problematic for the reconstruction in that it does not provide any kinematic resolution towards the Galactic centre and anti-centre directions. In the inner Galaxy, we know, however, that the gas also has radial motions due to the presence of the Galactic bar \citep{blitz1991}. The modelling of non-circular motions in this region is quite uncertain and we follow \citet{mertsch2021b} in considering two different gas flow models in order to somewhat bracket the systematic uncertainty. Consequently, a brief discussion for the two gas flow models is in order.

\begin{figure}[h!]
\centering
\includegraphics[scale=1]{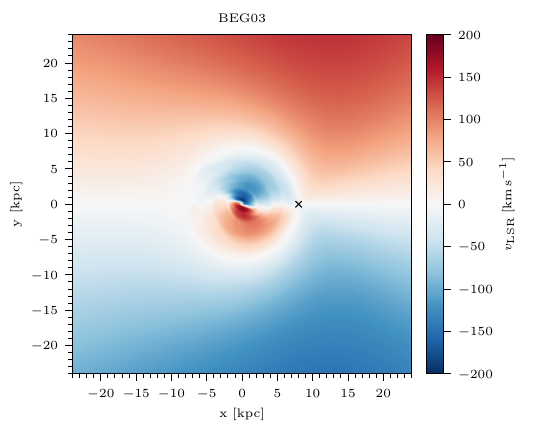}
\includegraphics[scale=1]{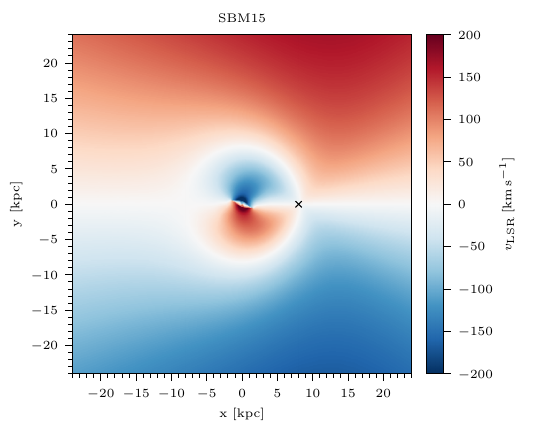}
\includegraphics[scale=1]{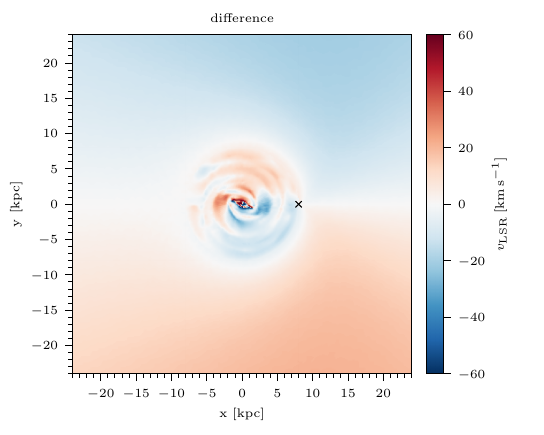}
\caption{Gas flow models adopted for the reconstruction. The radial velocity with respect to the local standard of rest velocity is shown as a function of position in the Galactic plane for the \citetalias{bissantz2003} (top) and \citetalias{sormani2015} (middle) models. The difference between the two models is shown in the bottom panel.}
\label{fig:velocity}
\end{figure}

The first model is based on the smoothed particle hydrodynamics simulation by \citet{bissantz2003} (hereafter \citetalias{bissantz2003}). The \citetalias{bissantz2003} velocity profile is shown in the top panel of Fig. \ref{fig:velocity}. We also note that this gas flow model was extended beyond $8 \, \text{kpc}$ using a flat rotation curve similar to the approach by \citet{pohl2008} \citep[see also][for more details]{mertsch2021b}.  

We also considered a modified version of the semi-analytic model for gas-carrying orbits in the potential dominated by the Galactic bar from \citet{sormani2015} (referred to as \citetalias{sormani2015}; see Appendix A of \citealt{mertsch2021b}). The \citetalias{sormani2015} velocity map seen by an observer moving with the velocity of the local standard of rest at a distance $8.15 \, \text{kpc}$ from the Galactic centre \cite{reid2019} and at an angle of $20^\degree$ with respect to the major axis of the bar is shown in the middle panel of Fig.~\ref{fig:velocity}. 

The difference between the \citetalias{bissantz2003} and \citetalias{sormani2015} models is shown in the bottom panel of Fig. \ref{fig:velocity}. We note that the two models do not agree very well within 2 kpc from the Galactic centre. Furthermore, there is no perturbation due to spiral structures in the \citetalias{sormani2015} model resulting in the difference from around $10 \, \text{km/s}$ to $30 \, \text{km/s}$ in certain regions in comparison to the \citetalias{bissantz2003} model. It should be noted that purely circular motion is assumed in both models beyond about 4 kpc from the Galactic centre and the differences as large as $30 \, \text{km/s}$ outside the solar circle are due to the different adopted rotation curves.
Differences between a velocity model and the true gas flow will manifest as gas placed at an incorrect distance along the line of sight, which is a concern common to any deprojection from the $\ell{}bv$ cubes alone. While either of our models likely exhibits deviations from the true gas flow, the differences between the gas densities that are reconstructed with either model can illustrate the dependence of features along the line of sight on the gas flow model to a certain degree.

\subsection{Bayesian inference}

The framework adopted here is similar to the one presented in \citet{mertsch2021b} for the reconstruction of the \Htwo{} density from the Galactic CO line survey.
We treated the reconstruction of the \HI{} density map as a high-dimensional Bayesian inference problem and sought the posterior distribution of the gas density for the given survey data using a method called Metric Gaussian Variational Inference (MGVI)~\citep{knollmuller2019}. The MGVI is a method from information field theory (IFT; \citealt[][]{IFT}) that relies on approximating the posterior distribution by a parametrised Gaussian distribution with the covariance taken to be the inverse Fisher information metric. The parameters of the approximate distribution are determined by repeatedly minimising the ``distance'' measured by the Kullback-Leibler divergence~\citep{kullback1968} between this distribution and the exact posterior. In practice, this is performed in an iterative fashion by alternating between computing the covariance for the current mean and updating the mean obtained from the minimisation of the KL divergence for the current covariance. The result is no explicit representation of the posterior distribution, but an ensemble of samples drawn from it that can be used to study the \HI{} density and its uncertainty.

\subsubsection{Data model}

The radio data of the 21 cm emission line provide the brightness temperature $T(\ell,b,v)$, and we would like to reconstruct the volume emissivity $\varepsilon(x,y,z)$. We assumed that we are always in the optically thin regime, rendering the relation linear, 
\begin{align}
T(\ell,b,v)&=R\left[\varepsilon\right](\ell,b,v)\n\\
&=\left.\displaystyle\int^{\infty}_0\df s ~\varepsilon(\vr)\delta(v-v_{\LSR}(\vr))\right|_{\vr=\vr(\ell,b,s)},\label{eq:R_ana}
\end{align}
where we also defined the response matrix $R=R^{xyz}_{lbv}$. The discretised version of the response matrix to transform from emissivity $\varepsilon_{\alpha\beta\gamma} \equiv \varepsilon(x_\alpha,y_\beta,z_\gamma)$ to brightness temperature $T_{ijk} \equiv T(\ell_i,b_j,v_k)$ is defined via
\begin{align}
R^{\alpha\beta\gamma}_{ijk}\varepsilon_{\alpha\beta\gamma}&=\int_{\ell_i}^{\ell_i+\Delta \ell}\df{l}\int_{b_j}^{b_j+\Delta b}\df b\n\\
&\times\int_{v_{k}}^{v_k+\Delta v}\df v\int^{s_{max}}_0\df s \frac{\varepsilon_{\alpha\beta\gamma}}{\Delta \ell\Delta b\Delta v}\n\\
&\!\!\!\! \times\left.\theta(\vr(\ell,b,s) \textrm{ in } r_{\alpha\beta\gamma})\delta\left(v-v_{\LSR}(\vr)\right)\right|_{\vr=\vr(\ell,b,s)} \! .
\end{align}

Peculiar motions can lead to deviations between the velocities described by the large-scale velocity field and the actual velocities. While this cannot be predicted, we can allow for some deviations by smearing the response matrix with a Gaussian kernel with variance $\sigma^2_v$,
\begin{eqnarray}
R'\left[\varepsilon_{xyz}\right]=\int\df \hat{v}\,\mathcal{G}(v-\hat{v})R\left[\varepsilon_{xyz}\right].
\end{eqnarray}
The velocity dispersion $\sigma_v$ has been estimated to be $9.2 \, \text{km/s}$ for the first quadrant ($0^\degree < l < 90^\degree$) and $9.0 \text{km/s}$ for the fourth quadrant ($270^\degree <l <  360^\degree$) \citep{malhotra1995}. We adopted $\sigma_v=10 \, \text{km/s}$ in line with \citet{nakanishi2003}. 

It should also be noted that the measured brightness temperature would also be altered by noise term which here we denote by $n$. We modelled $n$ as an additive Gaussian white noise $p(n)=\mathcal{G}(n,N)$ with the diagonal covariance in harmonic space $\tilde{N}=2\pi\sigma^2_n\delta(\mathbf{k}-\mathbf{k}')$ as in \citet{mertsch2021b}. We approximated the noise level of the HI4PI data \citep[see Tab. 1 in][]{HI4PI2016} with $\sigma_n=50 \, \text{mK}$. The final data model for the brightness temperature is then
\begin{eqnarray}
T_{lbv} = R'\left[\varepsilon_{xyz}\right] + n_{lbv} \, .
\end{eqnarray}
The likelihood function can then be written as (see also \citealt{mertsch2021b} for more details):
\begin{eqnarray}
p(T_{lbv}|\varepsilon_{xyz})=\mathcal{G}(T_{lbv}-R'[\varepsilon_{xyz}],N).
\end{eqnarray}

\subsubsection{Prior}
\label{sec:prior}

In order to infer the posterior distribution, we have to specify also the prior. We modelled the \HI{} emissitivity as a log-normal distributed random field meaning \mbox{$s(\vr)=\ln\left[\varepsilon(\vr)/\varepsilon_0\right]$} follows a normal distribution. We further assumed $s(\vr)$ to be statistically homogeneous and isotropic, such that the two-point correlation function in the harmonic domain could be characterised simply as $\langle \tilde{s}(\vk)\tilde{s}(\vk')\rangle=2\pi P(k) \delta(\vk-\vk')$, where $P(k)$ is the power spectrum. To allow for enough flexibility, we modelled $P(k)$ as the product of a (broken) power law and a random field (in $\log k$),
\begin{align}
\sqrt{P(k)} = \exp & \left[\vphantom{\dfrac{\frac{}{}}{\frac{}{}}} (\mu_y + \sigma_y \phi_y) + (\mu_m + \sigma_m \phi_m) \log(k) \right. \n\\
&\qquad\qquad\qquad \left. + \mathcal{F}^{-1} \!\! \left\{ \frac{a}{1 + t^2/t_0^2} \tau(t) \right\} \right] \! ,
\end{align}
Here, $\phi_y$ and $\phi_m$ represent the slope and the variance which are random variables following the Gaussian distributions $\mathcal{G}(\mu_y,\sigma_y)$ and $\mathcal{G}(\mu_m,\sigma_m)$ respectively, $\mathcal{F}^{-1}$ denotes the inverse Fourier transform from the variable $t$ to its conjugate $\log(k)$, and $\tau(t)$ is the above-mentioned random field (in $\log(k)$). The metaparameters were fixed as follows: $\mu_y=-13$, $\sigma_y=1$, $\mu_m=-4$, $\sigma_m=0.2$, $a=1$, and $t_0=0.1$. We reconstructed the random field $\tau(t)$ at the same time as the signal itself and, thus, determined $P(k)$ together with the gas density. Where the data were constraining enough, the data were able to deviate from the prior structure; where they were not the gas density was reconstructed with a realisation of the log-normal field prescribed by the power spectrum.

We note that the assumption of a homogeneous random field cannot be fully realised as the \HI{} is localised in a disk of finite extent in radius. Ignoring this can lead to biases, most severely affected by the quick fall-off of gas density with distance $z$ from the disk. We therefore scaled the signal field $\varepsilon_{xyz}$ with an exponential profile in $z$, $\exp\left(-|z|/z_{\text{h}}\right)$, where the scale-height $z_{\text{h}}$ varies with galactocentric radius $r$ as
\begin{eqnarray}
z_{\text{h}}=z^0_h\exp\left(\frac{r-R_\odot}{9.8 \,{\rm kpc}}\right) \, .
\end{eqnarray}
The scale of radial variation is the result of an earlier analysis~\citep{kalberla2008}, and we fixed $z^0_h = 150 \, \text{pc}$. While this $z$-profile does not preclude the possibility that other profiles are found by the reconstruction, we find that $z^0_h = 150 \, \text{pc}$ in particular leads to consistent results.

The inference method was implemented using the \texttt{NIFTy} package~\citep{nifty1,nifty3}\footnote{\url{https://gitlab.mpcdf.mpg.de/ift/nifty}}, specifically \texttt{NIFTy5}~\citep{nifty5}, which was previously applied to a variety of problems ranging from the reconstruction of the three-dimensional dust density in the Galaxy from reddening data~\citep{leike2019} to radio interferometry~\citep{arras2019}.

\subsubsection{Reconstructed density}

The conversion from the velocity-integrated brightness temperature $T_{B}$ to the \HI{} column density is normally defined~\citep{draine2011} through
\begin{eqnarray}
N_{\HI{}}(\ell,b)=X_{\HI{}}\int\df v~ T(\ell,b,v) \, , \label{eq:column}
\end{eqnarray}
where $N_{\HI{}}$ is the total column density for the line of sight $(\ell,b)$ and $X_{\HI{}}\simeq 1.823\times 10^{18}$ cm$^{-2}$ (K km s$^{-1}$)$^{-1}$. 

We note that Eq. \ref{eq:column} is applicable only in the optically thin limit which means that the derived column density strictly only provides a lower limit for the true column density in the disk \citep{nakanishi2003,johannesson2018}. The density can then be obtained from the reconstructed volume emissivity as
\begin{eqnarray}
n_{\HI{}}(\vr)=X_{\HI{}}\varepsilon(\vr).
\end{eqnarray}

\section{Results and discussion}
\label{sec:results}

The main results of our analysis are the three-dimensional density maps of Galactic \HI{} for the \citetalias{bissantz2003} and \citetalias{sormani2015} gas flow models. These maps are presented on a Cartesian grid with $(n_x,n_y,n_z)=(768,768,64)$ stretching from $-24 \, \text{kpc}$ to $24 \, \text{kpc}$ in the $x$- and $y$-directions and from $-2 \, \text{kpc}$ to $2 \, \text{kpc}$ in the $z$-direction. Therefore, the spatial resolution for these reconstructions is $62.5 \, \text{pc}$, that is the same as for the \Htwo{} reconstruction of \citet{mertsch2021b} and the highest resolution among all three-dimensional reconstructions of \HI{} line surveys to date. We have made our maps accessible on zenodo\footnotemark.

\footnotetext{\url{https://doi.org/10.5281/zenodo.5956696}}

\begin{figure*}[h!]
\centering
\includegraphics[width=6.5in]{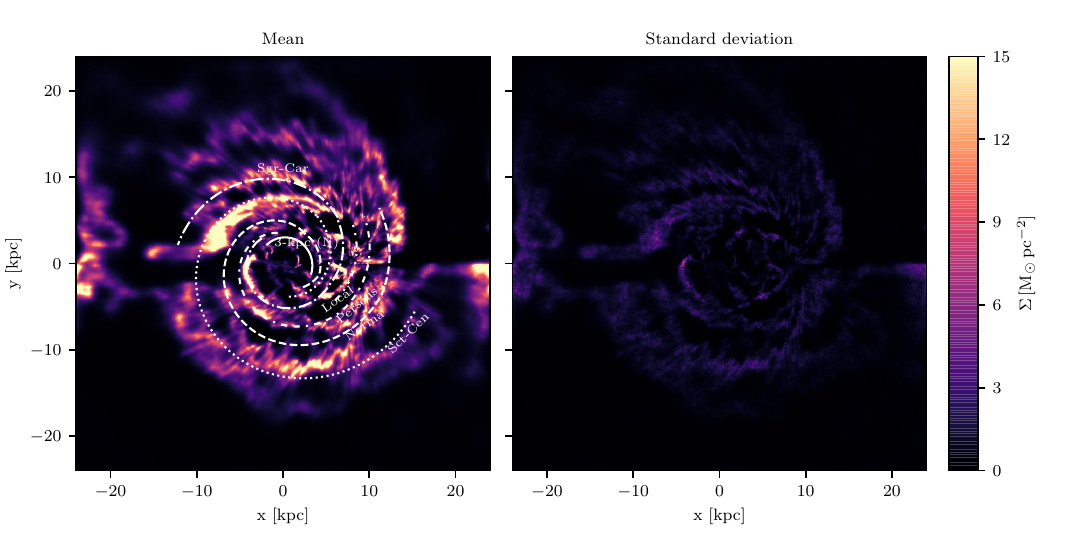}\\
\includegraphics[width=6.5in]{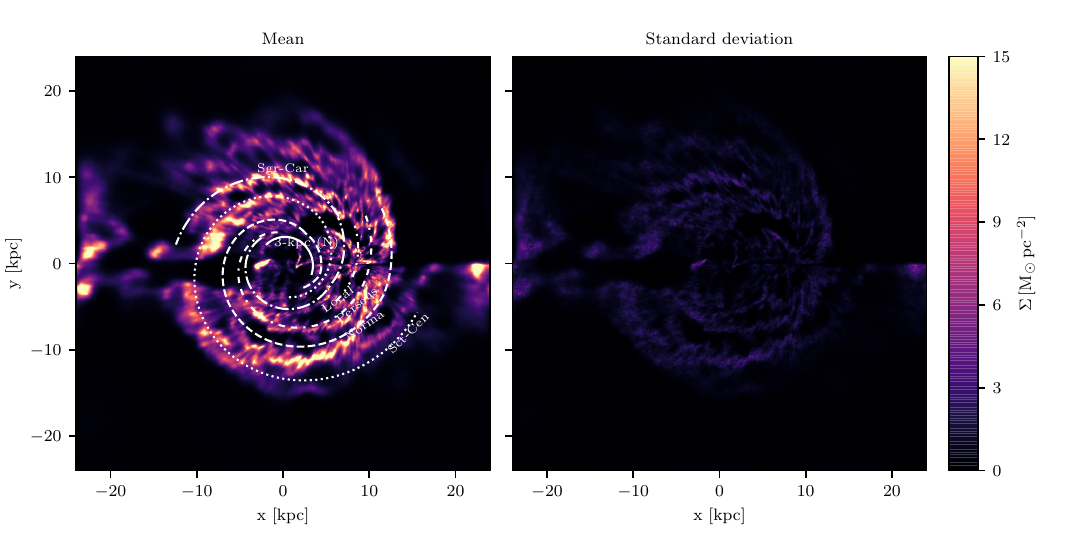}
\caption{Reconstructed mean of \HI{} surface mass number density for the gas flow model \citetalias{bissantz2003} (top panels) and \citetalias{sormani2015} (bottom panels).}
\label{fg:surfmassdens_mean_uncert}
\end{figure*}

We present in Fig.~\ref{fg:surfmassdens_mean_uncert} the mean and standard deviation of the \HI{} surface mass density for both the \citetalias{bissantz2003} (top panels) and \citetalias{sormani2015} (bottom panels) gas flow models. It is clear that the survey data are successfully reconstructed into localised clusters of gas following the spiral structures (fitted independently with data of molecular masers associated with very young high-mass stars) from \citet{reid2019}. Generally, the maps agree on the presence of large-scale features such as spiral arms, but their exact position and other properties differ between the gas flow models. This is expected given the disagreement between the radial velocities (see Fig.~\ref{fig:velocity}) which upon deprojection results in different distances to emission features. Some of these differences are also present in the reconstructions of \Htwo{}, and they have been linked to local extrema in the gas flow models \citep{mertsch2021b}. 

For example, for the \citetalias{bissantz2003} model there seems to be a spur-like feature stretching from $(x,y)\simeq (-8,2)\, \text{kpc}$ to $(x,y)=(-4,6)\, \text{kpc}$, and it is rather tempting to identify this feature with the Scutum-Centaurus (Sct-Cen) spiral arm. However, this is probably local emission with velocities close to $v_{\LSR} \approx 0 \, \text{km/s}$ that is misreconstructed at large distances from the Solar System: Brightness temperature spectra peak at $v\gtrsim 5 \, \text{km} \, \text{s}^{-1}$ for lines of sight with $-70^{\circ}\lesssim l \lesssim -20^{\circ}$. The gas flow models, however, give negative $v_{\LSR}$ inside the solar circle along these directions. Part of the emission is therefore placed outside the solar circle, while the large latitudinal spread indicates that the emission is local. Another noticeable difference that can be spotted is the small arc stretching from $(x,y)=(-5,0)\, \text{kpc}$ to $(x,y)=(-1,-4)\, \text{kpc}$ along the Sagittarius-Carina (Sgr-Car) arm in the \citetalias{bissantz2003} reconstructed map. This arc does not appear in the \citetalias{sormani2015} reconstructed map which again might be linked to the arc in the same position in the map showing the difference in velocity profiles (see the bottom panel of Fig. \ref{fig:velocity}). 

\begin{figure*}[p]
\centering
\includegraphics[scale=1]{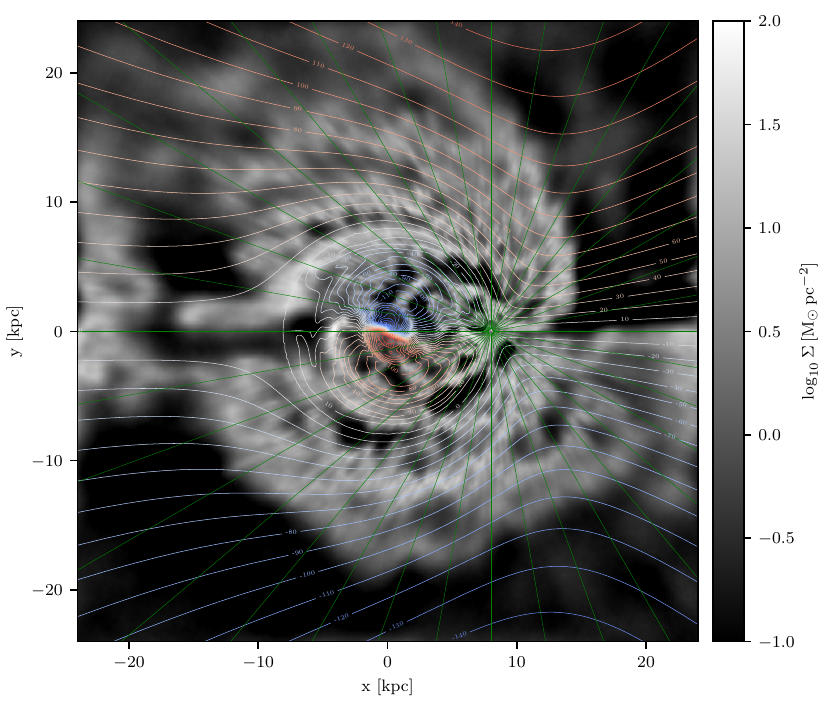} \\
\includegraphics[scale=1]{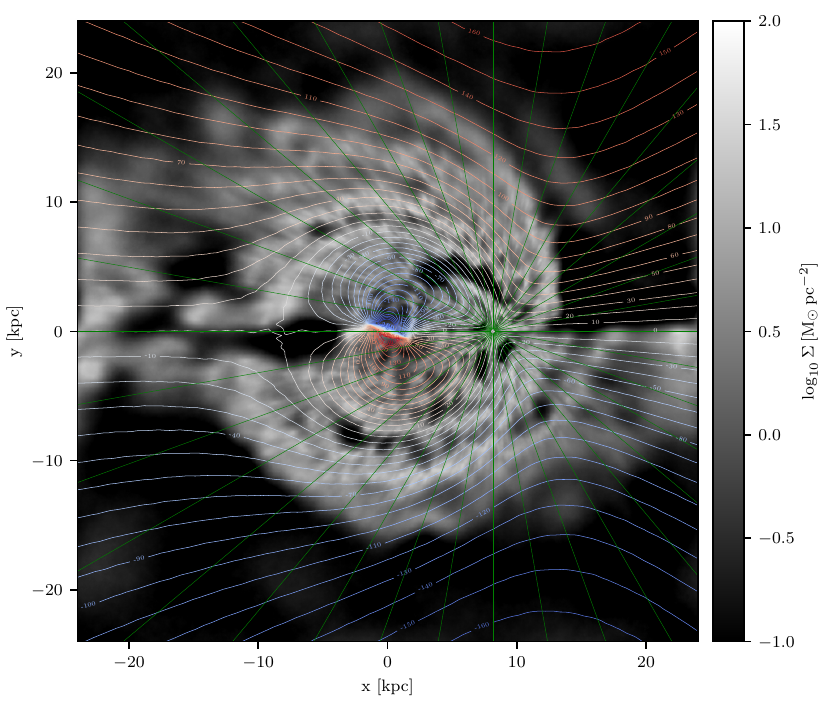}
\caption{Projected mean gas density $\Sigma$ on a logarithmic colour scale, overlaying both contours in $v_{\LSR}$ and a grid in longitude. \textbf{Top:} For the \citetalias{bissantz2003} gas flow model. \textbf{Bottom:} For the \citetalias{sormani2015} model.}
\label{fg:surfmassdens_vLSR}
\end{figure*}

Another version of the reconstructed mean density overlaid with velocity contours from the gas flow models and a longitude grid is shown in Fig.~\ref{fg:surfmassdens_vLSR}. Since the dynamical range of the gas density is rather large, as expected for a log-normal field, we have presented this map on a logarithmic colour scale to better highlight features that are not visible on a linear scale. For the reconstruction based on the \citetalias{bissantz2003} gas flow model, it is apparent that some of the bright features are aligned with regions of strong gradients in the velocity field, for example the arc stretching from $(x,y)=(-5,0)\, \text{kpc}$ to $(x,y)=(-1,-4)\, \text{kpc}$ that was already mentioned above. 

\begin{figure*}[p]
\centering
\includegraphics[scale=1]{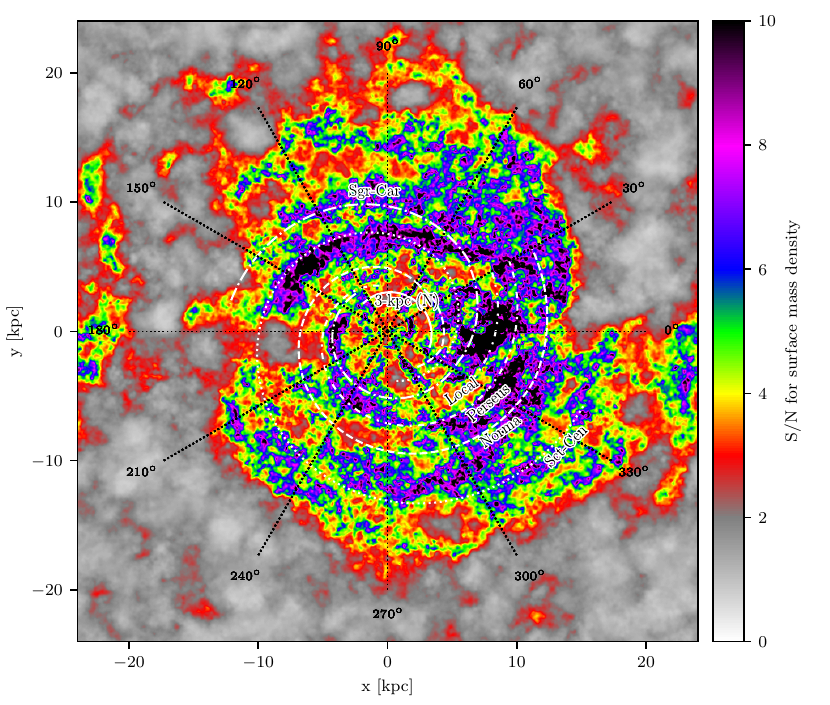} \\
\includegraphics[scale=1]{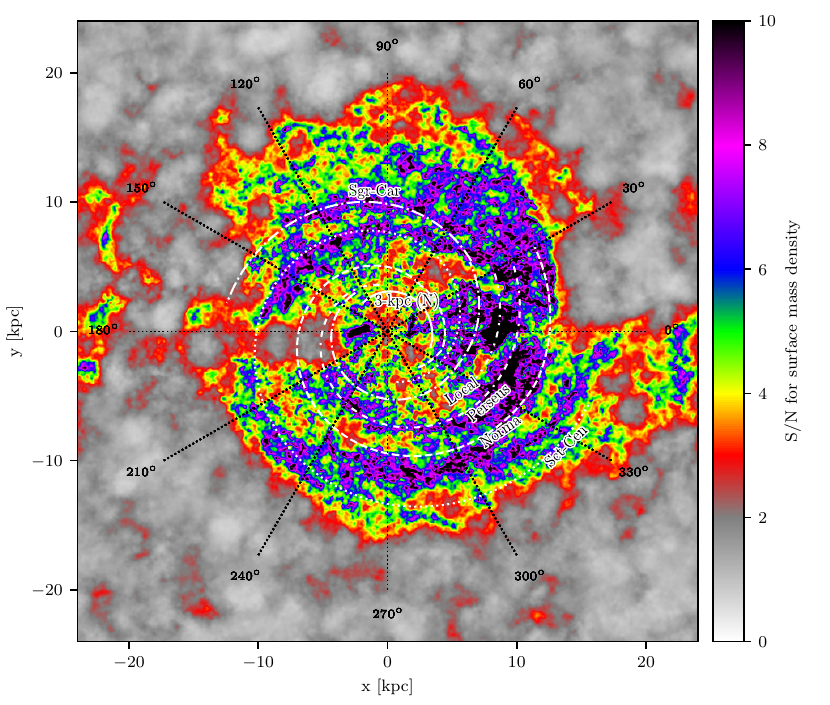}
\caption{S/N of the mean gas density, with the spiral arms of \citet{reid2019} overlaid (white lines) and a grid of galactocentric azimuth $\varphi$ (black lines). \textbf{Top:} For the \citetalias{bissantz2003} gas flow model. \textbf{Bottom:} For the \citetalias{sormani2015} model.}
\label{fg:surfmassdens_snr}
\end{figure*}

So far, we have discussed localised structures in the two models based on the reconstructed mean. However, some of the localised structures with relatively high surface mass density might also have rather large uncertainties. Unlike previous methods, the Bayesian inference method readily provides an estimate of the uncertainty. It might be more informative to judge the validity of specific features by comparing the mean $\mu$ with its uncertainty $\sigma$. We therefore followed \citet{mertsch2021b} and defined the signal-to-noise ratio (S/N) as $\mu/\sigma$. The surface mass density S/N overlaid with the spiral arms from \citet{reid2019} is presented in Fig.~\ref{fg:surfmassdens_snr} for the \citetalias{bissantz2003} (top panel) and \citetalias{sormani2015} (bottom panel). We shall now proceed with a more detailed discussion on some notable features of the reconstructed maps of the \citetalias{bissantz2003} and \citetalias{sormani2015} models:
\begin{itemize}
\item Most of the structures with a S/N roughly above 3 could be associated with spiral arms, especially for the Sct-Cen, Sgr-Car, Local and Perseus arms. We note, however, that little \HI{} has been reconstructed along the Norma arm in the inner Galaxy for either model (see Fig. \ref{fg:surfmassdens_mean_uncert}). 
\item Interestingly, even though significant structures with relatively high S/N are present around the Local arm in both maps in Fig.~\ref{fg:surfmassdens_snr}, the Local arm seems to be better reconstructed in the \citetalias{bissantz2003} model (see the top left panel of Fig. \ref{fg:surfmassdens_mean_uncert}). The corresponding structure in the \citetalias{sormani2015} reconstructed map seems to be more scattered around the Local arm. However, we caution again that this might be an artefact of the spiral structure already present in the velocity model for the case of the \citetalias{bissantz2003} model.
\item The \citetalias{bissantz2003} reconstructed map seems to extend to larger galactocentric radii than the \citetalias{sormani2015} one. Some clusters of \HI{} gas outside the solar circle are reconstructed at different distances in the two models due to the different rotation curves adopted beyond $4\, \text{kpc}$ from the Galactic centre. Because of the rather small velocity gradient, this easily translates into differences of the order of a kiloparsec and thus affects the association with spiral arms. We could see, for example, the arc of \HI{} gas in the \citetalias{bissantz2003} reconstructed map stretching between $(x,y)\simeq (10,15)\, \text{kpc}$ and $(x,y)=(12,2)\, \text{kpc}$, which is a little off the Norma arm (see the top left panel of Fig.~\ref{fg:surfmassdens_mean_uncert}). In the \citetalias{sormani2015} reconstructed map, this arc instead ends up on the Norma arm.
\item The total \HI{} masses reconstructed within $20\, \text{kpc}$ from the Galactic centre are \mbox{$M_{\HI{}}(r\leq 20 \, {\rm kpc})\simeq 3.42 \times 10^9 M_{\odot}$} for the \citetalias{bissantz2003} model and \mbox{$M_{\HI{}}(r\leq 20 \, {\rm kpc})\simeq 2.89 \times 10^9 M_{\odot}$} for the \citetalias{sormani2015} model. These values are comparable to the result from \citet{nakanishi2003}, who have found $M_{\rm HI}(r\leq 17 \, {\rm kpc})\simeq 2.5\times 10^9 M_{\odot}$.
\end{itemize}

\begin{figure*}[h!]
\includegraphics[scale=1]{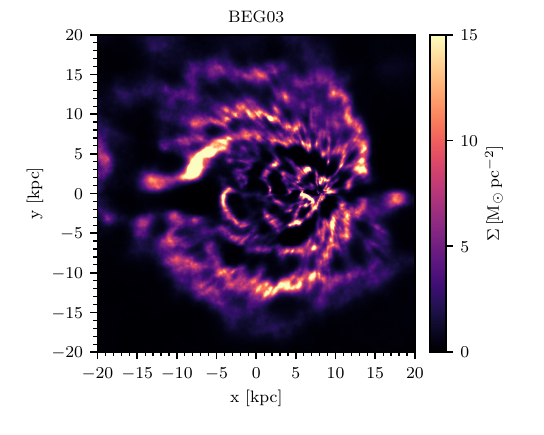} \includegraphics[scale=1]{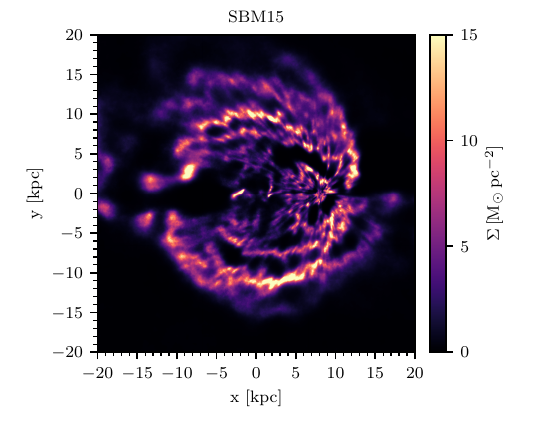} \\
\includegraphics[scale=1]{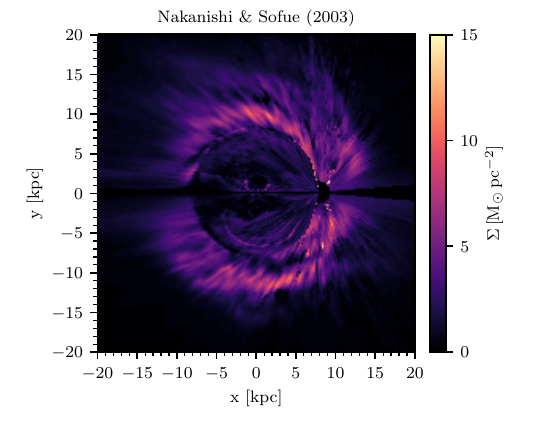}
\includegraphics[scale=1]{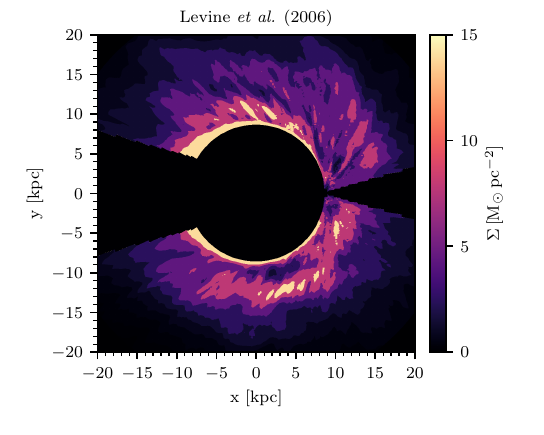}\\
\includegraphics[scale=1]{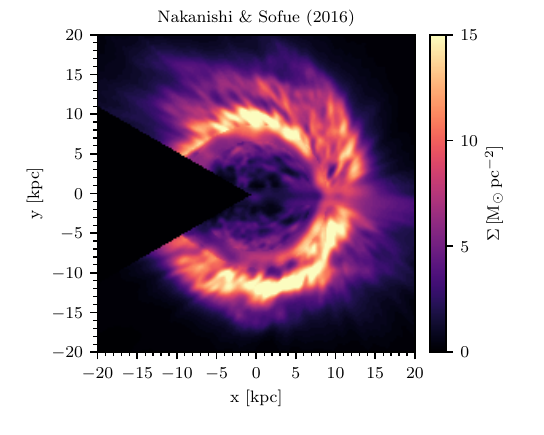}
\includegraphics[scale=1]{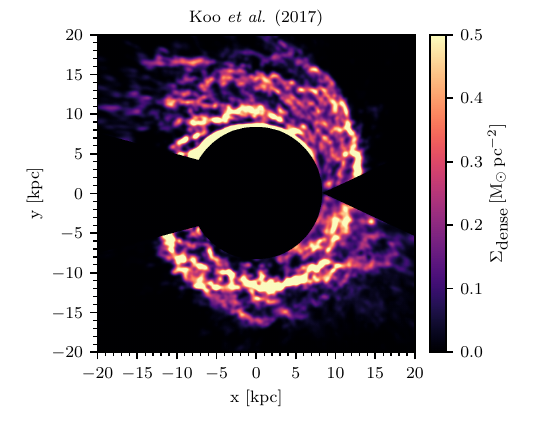}
\caption{Comparison of both our gas surface density construtions (top left and top right) with those of \citet{nakanishi2003} (middle left), \citet{levine2006a} (middle right), \citet{nakanishi2016} (bottom left), and \citet{koo2017} (bottom right).}
\label{fg:comparison}
\end{figure*}

We note that the comparison between the surface mass densities reconstructed with both gas models might not reflect the full systematic uncertainty due to our ignorance of the true velocity field. One possible way of estimating deviations between the gas flow models and the true velocity field makes use of the velocities measured for molecular masers in certain high-mass star-forming regions~\citep{reid2019}. While deviations from a circular rotation model are clustered around zero, the spread can be several tens of $\text{km} \, \text{s}^{-1}$~\citep{tchernyshyov2017,tchernyshyov2018} (see also Fig.~6 of \citealt{reid2019}).

It has also been suggested that when gas spectra were combined with differential dust extinction data, the Galactic velocity field can be reconstructed~\citep{tchernyshyov2017}. When this is applied, localised devitations of the reconstructed velocities from a circular rotation curve as large as $30 \, \text{km} \, \text{s}^{-1}$ have been inferred. As stressed in Sect.~\ref{sec:velocity}, these devitations can lead to gas placed at incorrect distances along the line of sight, in particular, along directions of small velocity gradients. While these studies are very promising, they require a rather tight correlation between gas and dust, that is, essentially a constant gas-to-dust ratio. We have turned this problem around and compared our reconstructed gas maps with gas maps converted from dust maps assuming this constant gas-to-dust ratio. In Appendix~\ref{appendixA}, we present the results and show that some features in our gas maps can be clearly identified with features in extinction maps while others cannot.

In Fig. \ref{fg:comparison}, we again show the surface mass density in the \citetalias{bissantz2003} and \citetalias{sormani2015} models together with previous reconstructions from \citet{nakanishi2003} (middle left), \citet{levine2006a} (middle right), \citet{nakanishi2016} (bottom left), and \citet{koo2017} (bottom right). All surface density maps are presented with the same dynamical range, except for the one by \citet{koo2017}. This is because \citet{koo2017} focused essentially on regions of dense \HI{} concentrations and, thus, provided the surface mass density map only for the densest structures. A few comments on the similarities and differences among our reconstructions and previous ones are in order:
\begin{itemize}
\item Both the \citetalias{bissantz2003} and \citetalias{sormani2015} reconstructed maps have spiral structures beyond the solar circle that are similar to all earlier analyses. For example, segments of the Sgr-Car arm in the region with $y>0$ or segments of the Sct–Cen arm appear in all of these reconstructed maps. We note, however, that not many localised structure with high surface mass density being reconstructed inside the solar circle for the maps of \citet{nakanishi2003} and \citet{nakanishi2016}. In our case, we clearly see more high surface mass-density clusters of gas inside the solar circle. 
\item As discussed above, peculiar motion of the gas can result in the finger-of-god effect, which manifests itself as elongated structures along the lines of sight in most of the previous reconstructed maps (see e.g. \citealt{nakanishi2003,levine2006b,nakanishi2016}). In the Bayesian framework, these structures seem less apparent (see the top panels of Fig. \ref{fg:comparison}) thanks to both the consideration of correlations and the treatment of velocity dispersion in the response matrix.
\item Finally, our reconstructed maps show clusters of \HI{} gas in regions with galactocentric radii between $15$ and $20 \, \text{kpc}$ along the Galactic centre and anti-centre directions that were excluded in all previous studies. However, we caution that most of them have much lower S/N than other features, and their distances are, given the lack of association with known spiral structure, most likely unreliable.
\end{itemize}

\begin{figure*}[p]
\centering
\includegraphics[scale=1]{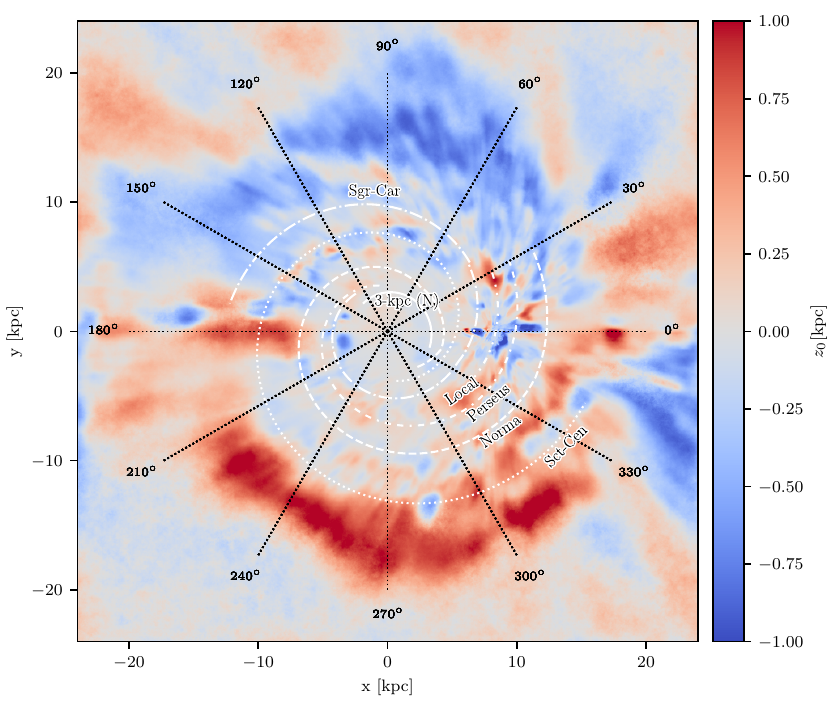}\\ \includegraphics[scale=1]{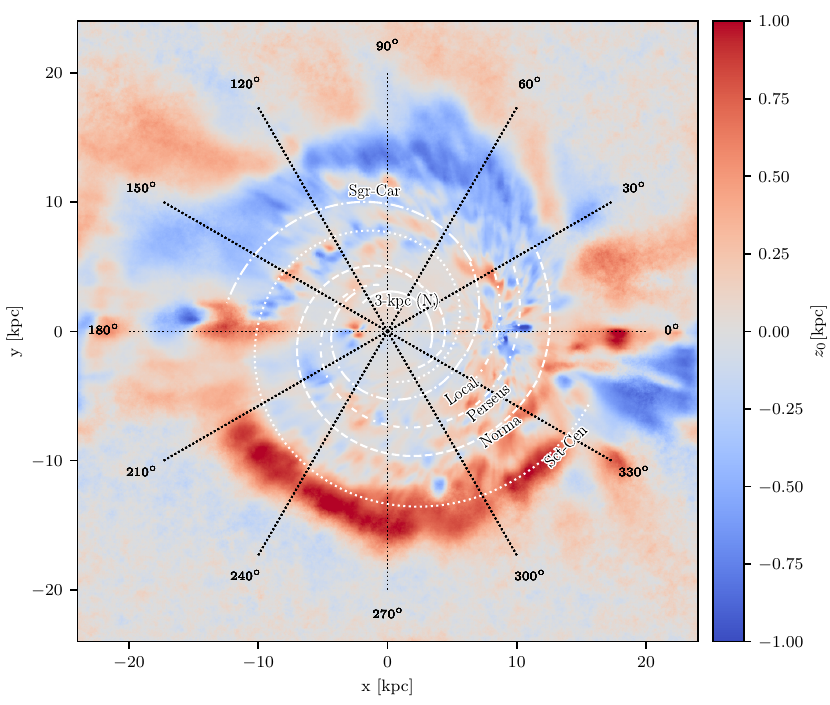}
\caption{Weighted mean of the \HI{} disk vertical offset (introduced in Eq. \ref{eq:z0}) from the reconstructed gas number density at for the \citetalias{bissantz2003} (upper panel) and \citetalias{sormani2015} (lower panel) models.}
\label{fg:warp}
\end{figure*}

\begin{figure}[h!]
\includegraphics[scale=1]{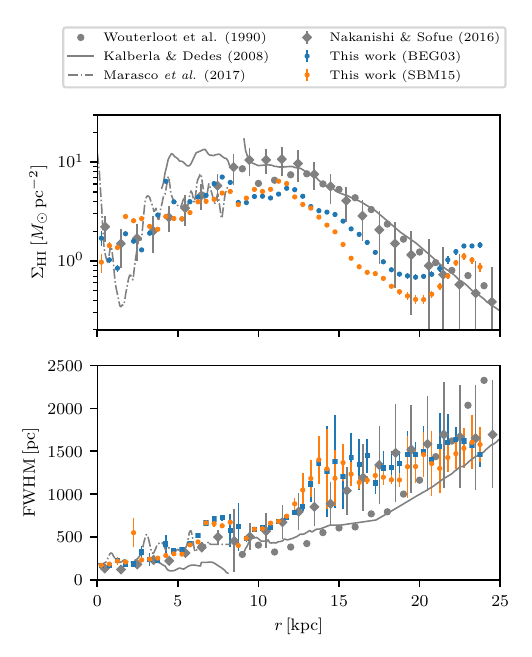}
\caption{Radial profiles of the surface mass density $\Sigma_{\HI{}}$ (upper panel) and FWHM of the $z$-profile (lower panel). We compare the results for our two gas flow models (\citetalias{bissantz2003} and \citetalias{sormani2015}) with the results of earlier analyses by \citet{wouterloot1990,kalberla2008,nakanishi2016,marasco2017}.}
\label{fig:profiles}
\end{figure}

We now examine the \HI{} profile in the $z$-direction. We present in Fig. \ref{fg:warp} the weighted mean of the \HI{} disk vertical offset defined as \citep{johannesson2018}
\begin{eqnarray}
z_0(x,y) = \frac{\int \df z \, z \, n_{\HI{}}(x,y,z)}{\int \df z \, n_{\HI{}}(x,y,z)} \, ,
\label{eq:z0}
\end{eqnarray}
where we approximated the integrals with sums over the bins in $z$-direction. It is clear that the midplane of the \HI{} disk is displaced from $z=0$ for galactocentric radii $r \gtrsim 12 \, \text{kpc}$ for both the \citetalias{bissantz2003} (upper panel) and \citetalias{sormani2015} (lower panel) models. This feature is sometimes referred to as the Galactic warp and has also been found in many of the previous studies \citep[see e.g.][]{burton1986,diplas1991,nakanishi2003,levine2006a,nakanishi2016,johannesson2018}. In our case, the \HI{} disk beyond $12 \, \text{kpc}$ from the Galactic centre bends towards negative $z$ in the region between $\varphi\simeq30^\circ$ to $\varphi\simeq150^\circ$ and positive $z$ in the region between $\varphi\simeq210^\circ$ to $\varphi\simeq330^\circ$ for both velocity models. We note, however, that the exact values of $z_0$ are slightly different between the two models in certain parts of the Galaxy. For example, the outer disk ($r > 12\, \text{kpc}$) seems to be more warped in the \citetalias{bissantz2003} map than in the \citetalias{sormani2015} map in the regions between $\varphi=60^\circ$ and $\varphi=90^\circ$ and between $\varphi=300^\circ$ and $\varphi=330^\circ$.

Some summary statistics of the \HI{} gas distribution in galactocentric radius can also be extracted from our reconstructed maps. We present in the upper panel of Fig.~\ref{fig:profiles} the surface mass density averaged in galactocentric rings $r_i \leq r < r_{i+1}$ where $r_i = i \Delta r$ with $\Delta r = 0.5 \, \text{kpc}$. In the \citetalias{sormani2015} model, the reconstructed \HI{} surface mass density increases slightly from \mbox{$\Sigma_{\HI{}}\simeq 1\,M_\odot \, \text{pc}^{-2}$} around the Galactic centre to the maximum value of \mbox{$\Sigma_{\HI{}}\simeq 6 \, M_\odot \, \text{pc}^{-2}$} at $r \simeq 12 \, \text{kpc}$. Similarly, the \HI{} surface mass density from the \citetalias{bissantz2003} model also has \mbox{$\Sigma_{\HI{}} \simeq 1 \, M_\odot \, \text{pc}^{-2}$} around the Galactic centre and the maximum value is \mbox{$\Sigma_{\HI{}} \simeq 7\, M_\odot \, \text{pc}^{-2}$}. The \citetalias{bissantz2003} profile, however, seems to have a few more local maxima and the \HI{} surface mass density is slightly higher than that of \citetalias{sormani2015} for rings between $r = 4 \, \text{kpc}$ and $r=8 \, \text{kpc}$. We note that the surface mass density with these galactocentric radii appears to be compatible with data from \citet{wouterloot1990} (filled circles in Fig.~\ref{fig:profiles}) and also comparable to the commonly quoted value of $5\, M_\odot \, \text{pc}^{-2}$ for the inner Galaxy \citep{dickey1990,kalberla2009}. Beyond $r \simeq 12\, \text{kpc}$, both models exhibit a continuously decreasing \HI{} surface mass density up to $r\simeq 20\, \text{kpc}$, with values slightly lower than data from \citet{wouterloot1990} and the fit (for Galacto-centric radii between $12.5$ and $30\, \text{kpc}$) by \citet{kalberla2008}. Finally, it is worth mentioning that, in the region $r\gtrsim 20\, \text{kpc}$, the \HI{} surface mass density again increases slightly, but we caution that this is likely due to the misreconstruction of gas in the Galactic centre and anti-centre directions.

In the bottom panel of Fig.~\ref{fig:profiles}, we show the spread of the reconstructed \HI{} disk around the midplane as a function of galactocentric radius $r$. Specifically, we fitted a Gaussian profile, as is oftentimes assumed, and display its full width at half maximum (FWHM), which is related to the standard deviation $\sigma$ as \mbox{$\text{FWHM} = 2 \sqrt{2 \ln 2} \sigma \simeq 2.4 \sigma$}. The warping means that the maximum position of the vertical profile in $z$ varies with $x$ and $y$, therefore, we fit the Gaussian profile for limited ranges of galactocentric radius $r$ and azimuth $\varphi$ and then averaged over the bins in $\varphi$ for each $r$ bin. The width increases steadily from the inner to the outer Galaxy. Unlike earlier work~\citep{wouterloot1990,kalberla2008}, however, we do not find an exponential behaviour and instead the FWHM almost saturates beyond $r \sim 15 \, \text{kpc}$. We stress that while some assumption on the gas being confined to the plane was necessary in order to model the gas density as a homogeneous random field, see Sect.~\ref{sec:prior}, the reconstruction algorithm has the freedom to deviate from the specific profile assumed if this is preferred by the data. We note that the reconstructed widths beyond $r \sim 15 \, \text{kpc}$ differ significantly between the different samples from the posterior distribution, which is reflected in the relatively large error bars. 

\section{Summary and perspectives}
\label{sec:summary}

We have presented new three-dimensional \HI{} maps with a spatial resolution of $62.5 \, \text{pc}$ deprojected from the HI4PI 21 cm line survey~\citep{HI4PI2016}. These maps were reconstructed using two gas flow models, namely the \citetalias{bissantz2003} and \citetalias{sormani2015} models. Both of them consider the effects of the Galactic bar. The former was derived from a simulation of the gas flow in a predetermined potential \citep{bissantz2003} and the latter is based on a model for gas-carrying orbits in the bar potential \citep{sormani2015}.

We assumed the optically thin limit such that the velocity-integrated brightness temperature and the \HI{} gas column density could be related by a simple relation. The deprojection was then carried out using MGVI~\citep{knollmuller2019}. This Bayesian framework has previously been adopted for the CO line survey in \citet{mertsch2021b} and allows evaluating not only the mean gas density but also its uncertainty. We can therefore distinguish between statistically significant structures and noise artefacts. While we have assumed the existence of correlations in configuration space, the specific form of the power spectrum was not imposed but instead determined together with the gas density during the reconstruction. 

We have made our mean \HI{} maps and their uncertainty available\footnotemark[\value{footnote}]. We examined the surface mass density of deprojected maps and found that there are structures that are compatible with spiral arms fitted from parallax measurements of masers in \citet{reid2019}. Some radial profiles out of the three-dimensional gas distribution were also investigated and apparently agree well with earlier analyses for regions within roughly $12\, \text{kpc}$ from the Galactic centre. The warping of the \HI{} disk, which has been identified in some of the previous studies, is observed in these reconstructions as well. 

In the future, we plan to improve this analysis in two main ways: First, we will relax the optically thin assumption, which will provide more reliable reconstructions of the \HI{} density, in particular, in the inner Galaxy. Second, we will use more flexible velocity models than those used here. In particular, we will constrain these models with the velocity information from molecular masers~\cite{reid2019}. If this is successful, even a joint reconstruction of gas density and velocity field could be attempted, as suggested by \citet{mertsch2021b}.

In addition, we note that dust reddening data might provide additional constraints for the three-dimensional distribution of gas. In fact, the IFT framework has been successfully applied to derive a three-dimensional view of nearby dust~\citep{leike2019,leike2022}, and results agree rather well with other analyses within $\sim 300$ pc around the Solar System~\citep{lallement2018,green2019}. If the correlation between dust and gas was known with sufficient accuracy, dust reddening data could be used as an additional input for the gas reconstruction. Moreover, by combining dust data with emission lines of HI and CO, a three-dimensional model for the line-of-sight velocity could be derived. \citet{tchernyshyov2017}, however, pointed out that a combined analysis requires matching current reddening data, which are local (within a few kiloparsecs from the Solar System), with emission line data, which presumably are global (the intensity is contributed by all the gas along the line of sight). In addition, the conversion between the dust and gas distributions is normally carried out by assuming a relation between the dust reddening and the amount of dust and a gas-to-dust mass ratio which might vary at different positions in our Galaxy~\citep{jenkins2009,peek2013,planck2015}. More importantly, the three-dimensional dust distribution in recent analyses seem to disagree at large heliocentric distances due to the uncertainty in distance estimates to nearby stars (see e.g. \citealt{green2019,leike2022}). Nevertheless, three-dimensional dust maps can offer important additional constraints for the gas reconstruction on kiloparsec scales, yet a convergence of different dust reconstructions seems desirable. We refer to Appendix \ref{appendixA} for more in-depth discussions of the potential difficulties in using different analyses of dust reddening data for the dust-to-gas conversion.

In the end, despite all the shortcomings, it is clear that the three-dimensional \HI{} maps we have provided here are improving over previous analyses and should have interesting implications for other research directions. For instance, combining these maps with recently reconstructed \Htwo{} maps from \citet{mertsch2021b}, the modelling of diffuse gamma-ray emission from cosmic-ray interactions with the Galactic disk (see e.g. \citealt{dundovic2021}) can be reevaluated and its uncertainty can be estimated.

\begin{acknowledgement}
We thank the referee, Dr. Joshua Peek for useful comments that helped improve the quality of the manuscript. We would like to thank also Gordian Edenhofer, Torsten Enßlin, Ralf Kissmann, and Andrés Ramírez for fruitful discussions.
\end{acknowledgement}

\bibliographystyle{aa}
\bibliography{mybib}

\begin{appendix}
\section{Deriving gas maps from dust reddening data}
\label{appendixA}

\begin{figure*}[h!]
\centering
\includegraphics[width=7.2in, height=3.4in]{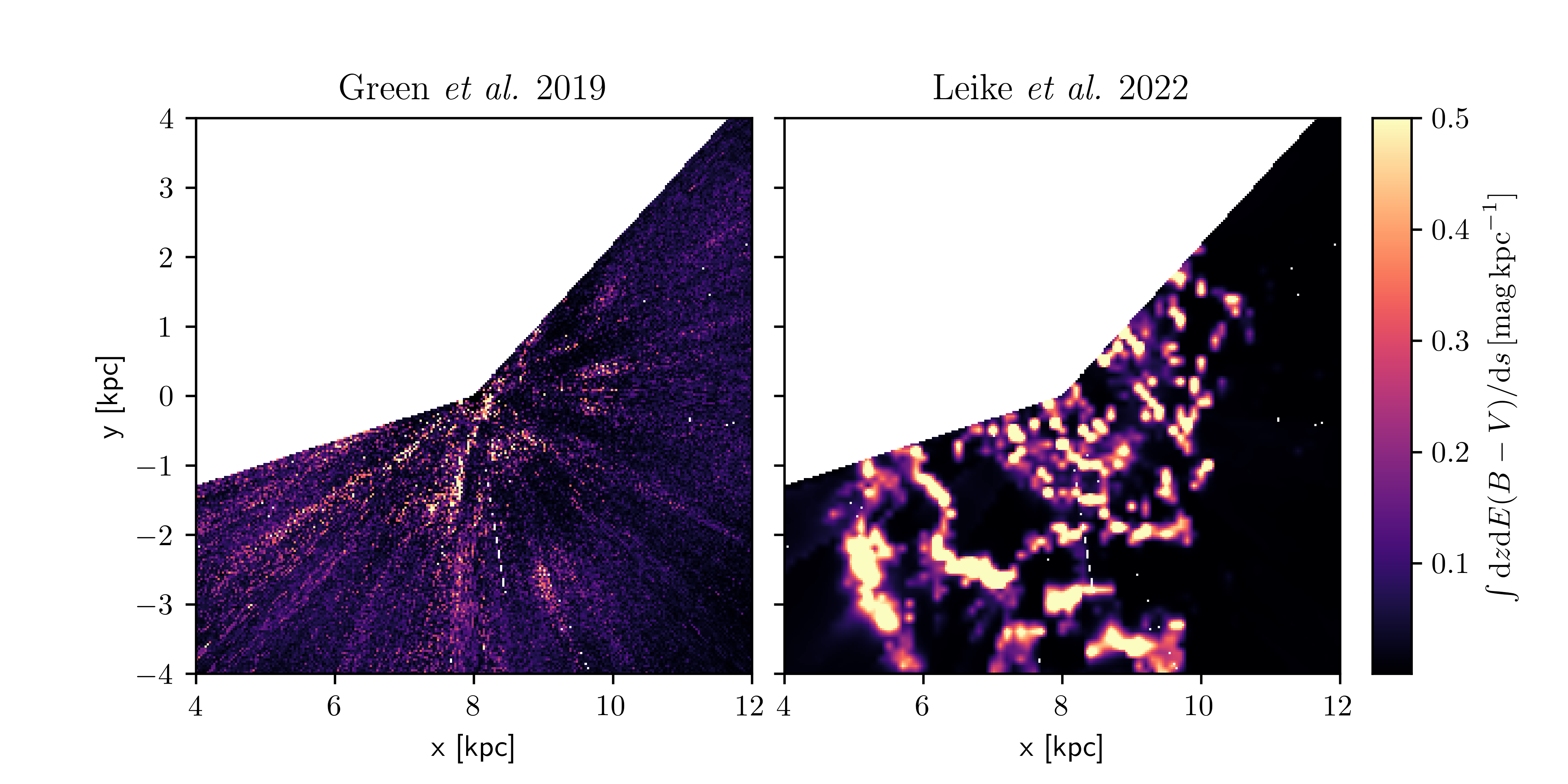}
\caption{Differential dust reddening integrated along the direction perpendicular to the Galactic disk $\int^{\infty}_{-\infty}{\rm d}z\, {\rm d}E(B-V)/{\rm d}s$ derived from the analyses of \citet{leike2022} (left) and \citet{green2019} (right).}
\label{fg:dust-map}
\end{figure*}

In this appendix, we first briefly discuss the differences in three-dimensional dust reddening data from the analyses by \citet{green2019} and \citet{leike2022}. We then derive the corresponding gas maps from these data sets and compare them to our reconstructions. 

We focus essentially on dust reddening data presented in recent works by \citet{green2019} and \citet{leike2022}, which provide three-dimensional maps of the differential dust reddening $\df E(B-V)/\df s$ and the differential dust extinction of $G$-band photons $\df A_{G}/\df s$ ($s$ denotes the heliocentric distance), respectively. In order to compare the two data sets, we converted the dust reddening into the $G$-band extinction assuming $E(B-V)\simeq A_V/3.1 \simeq 1.2 A_G/3.1$ (as in \citealt{leike2021}).

It was shown in \citet{leike2019} (an earlier version of \citealt{leike2022}) that, despite having slightly different inputs, these data sets agree well within roughly $300$ pc from the Solar System. This region is much smaller than the region on which our reconstructed gas maps are obtained. On larger scales, the similarities between different analyses might no longer be obvious. In Fig. \ref{fg:dust-map}, we present the differential dust reddening ${\rm d}E(B-V)/{\rm d}s$ integrated along the direction perpendicular to the Galactic disk for the analyses of \citet{green2019} (left) and \citet{leike2022} (right). The coordinate system adopted throughout this appendix was chosen to be the same as in the main text. It is clear that the two maps do not agree in general as \citet{leike2022} seems to give more clumpy structures, while structures in \citet{green2019} seem to be more elongated along the line of sight. More importantly, since $\int^{\infty}_{-\infty}{\rm d}z\, {\rm d}E(B-V)/{\rm d}s$ should provide the rough estimate for the surface density of dust, we could see that the surface density of dust in the case of \citet{green2019} decreases at large heliocentric distance. More investigations might be required to better understand the differences between these analyses. We caution that given the current disagreement in the dust reddening maps, it might not be straightforward to adopt these data sets as constraints for gas models.

\begin{figure*}[h!]
\centering
\includegraphics[width=3.2in, height=2.7in]{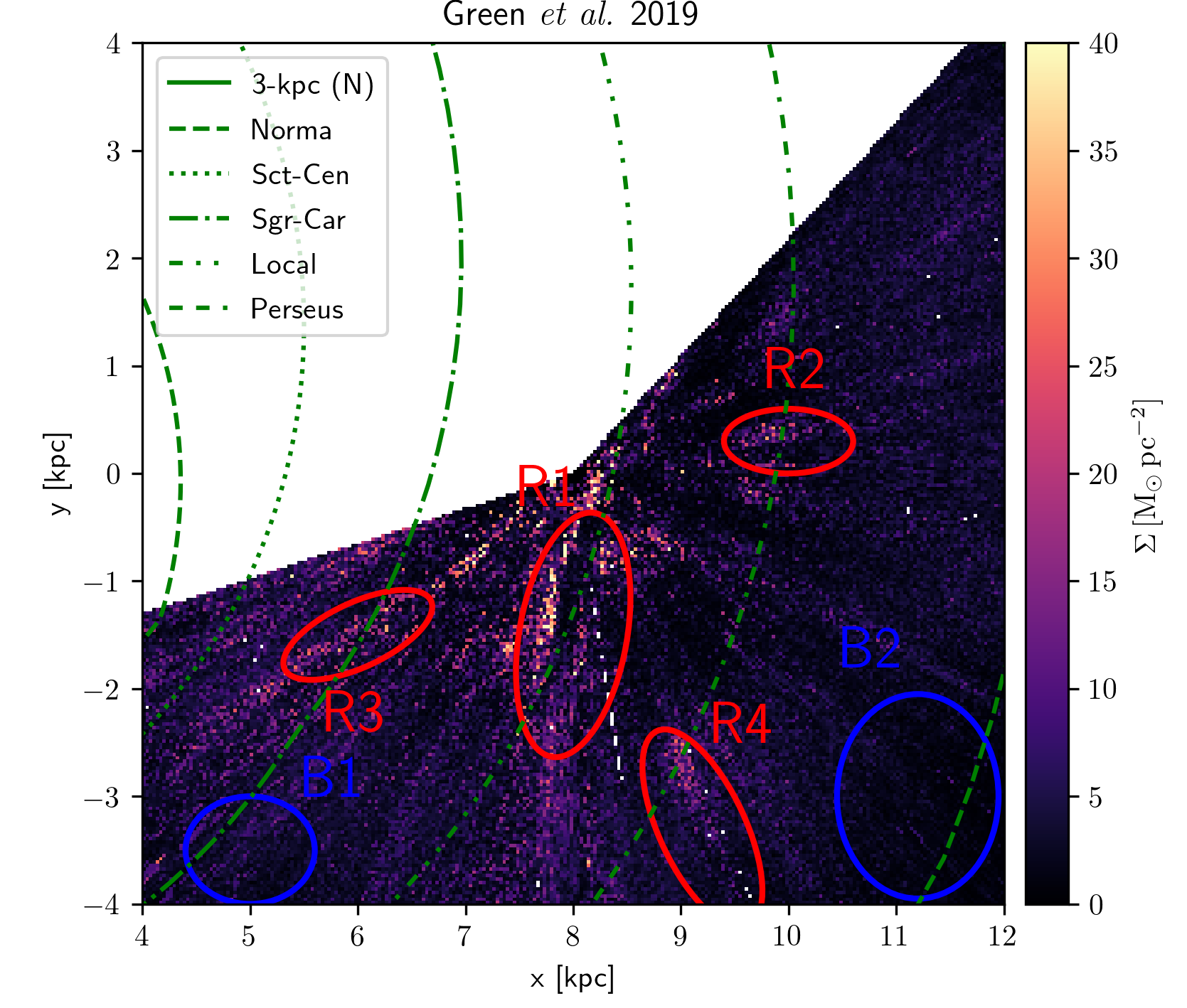}
\includegraphics[width=3.2in, height=2.7in]{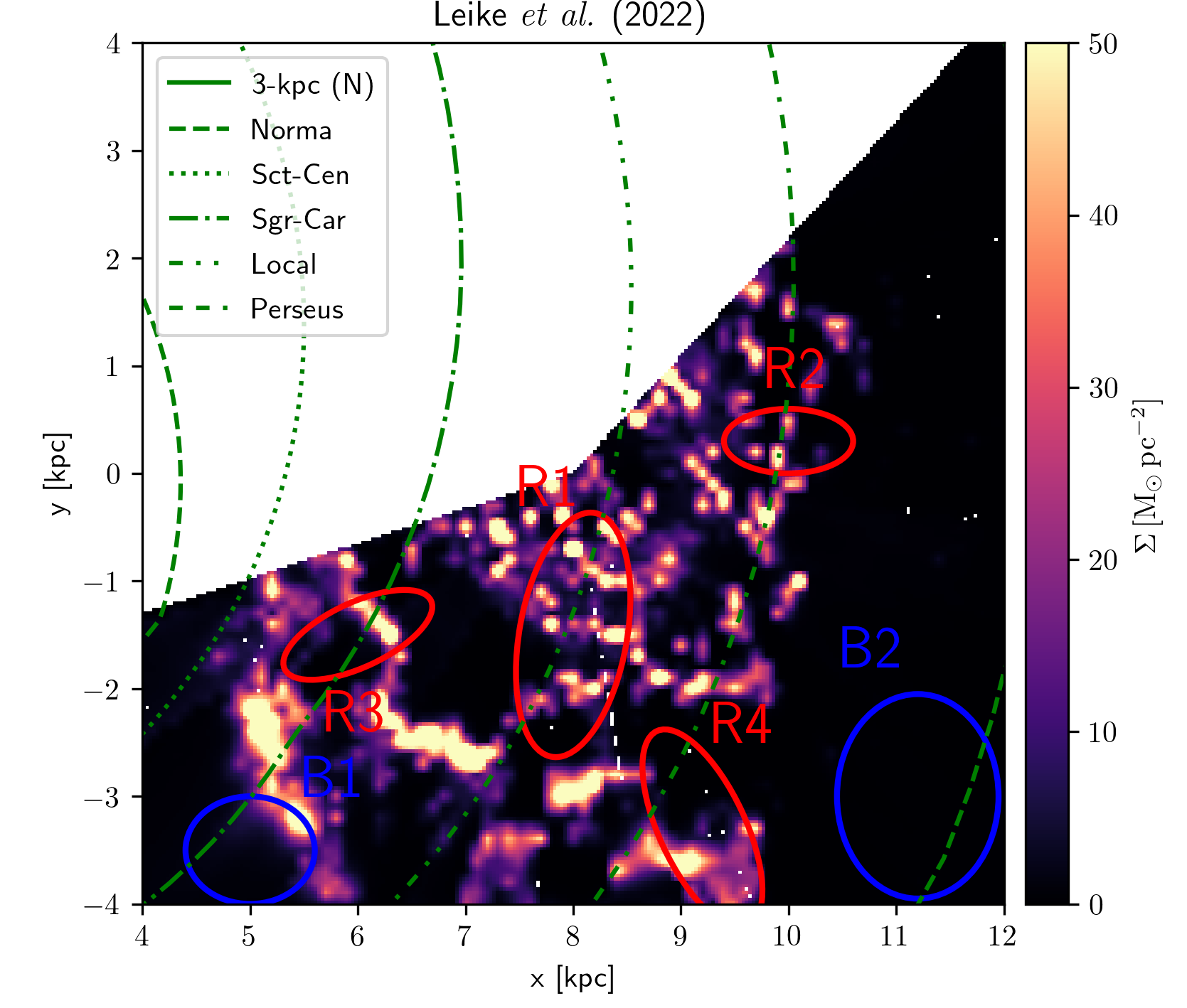}\vspace{0.5cm}\\
\includegraphics[width=3.2in, height=2.7in]{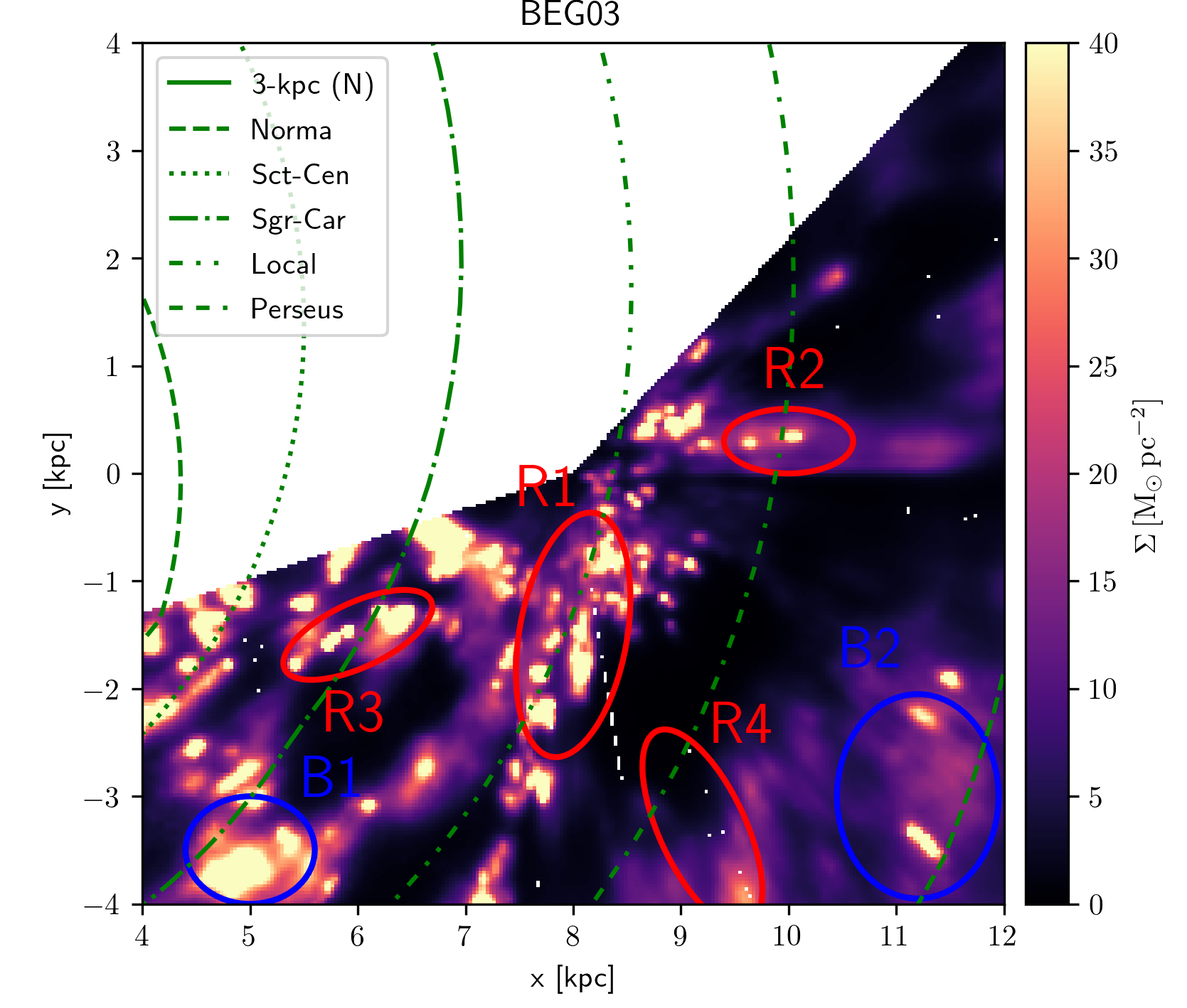}
\includegraphics[width=3.2in, height=2.7in]{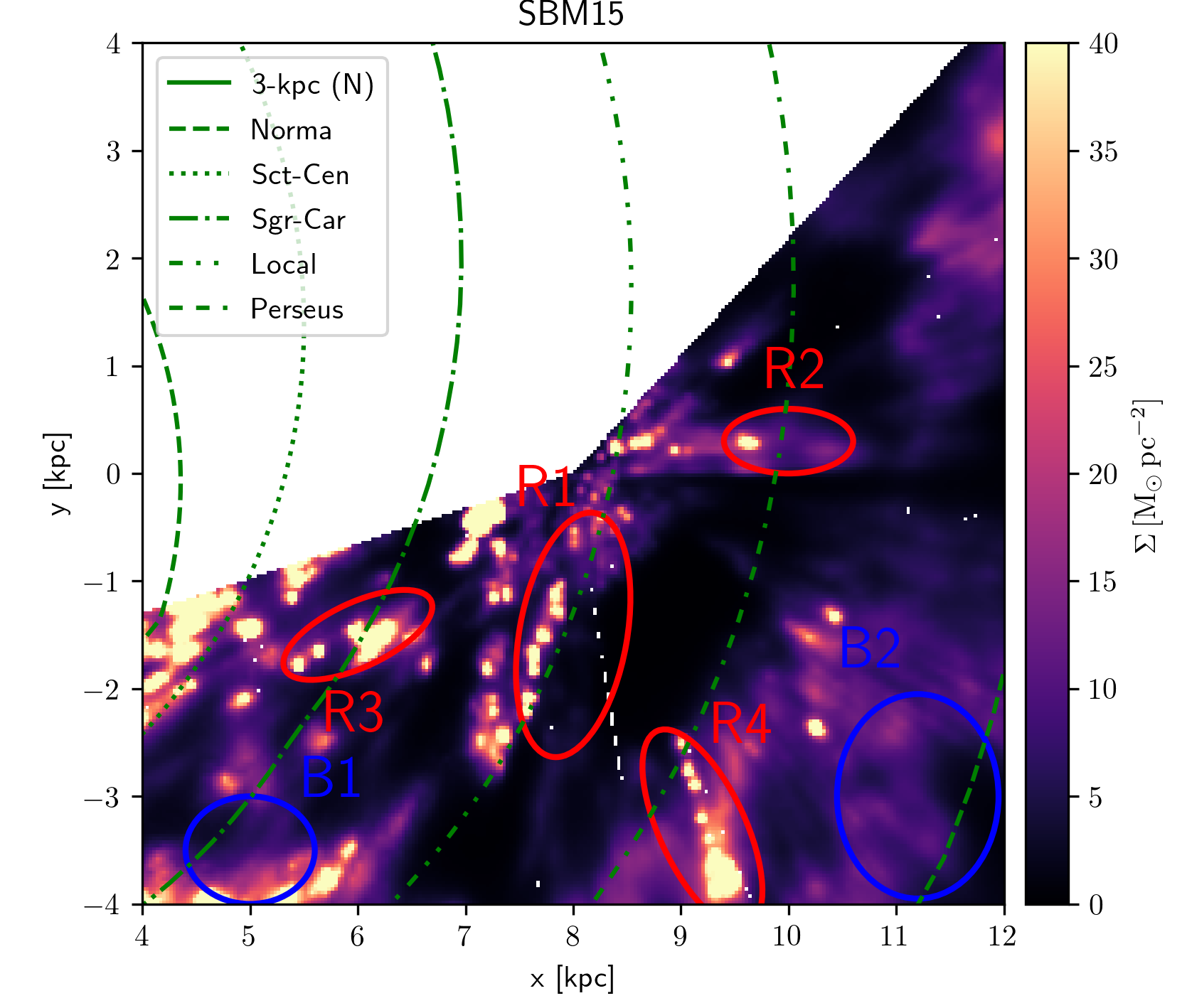}\vspace{0.5cm}\\
\includegraphics[width=3.2in, height=2.7in]{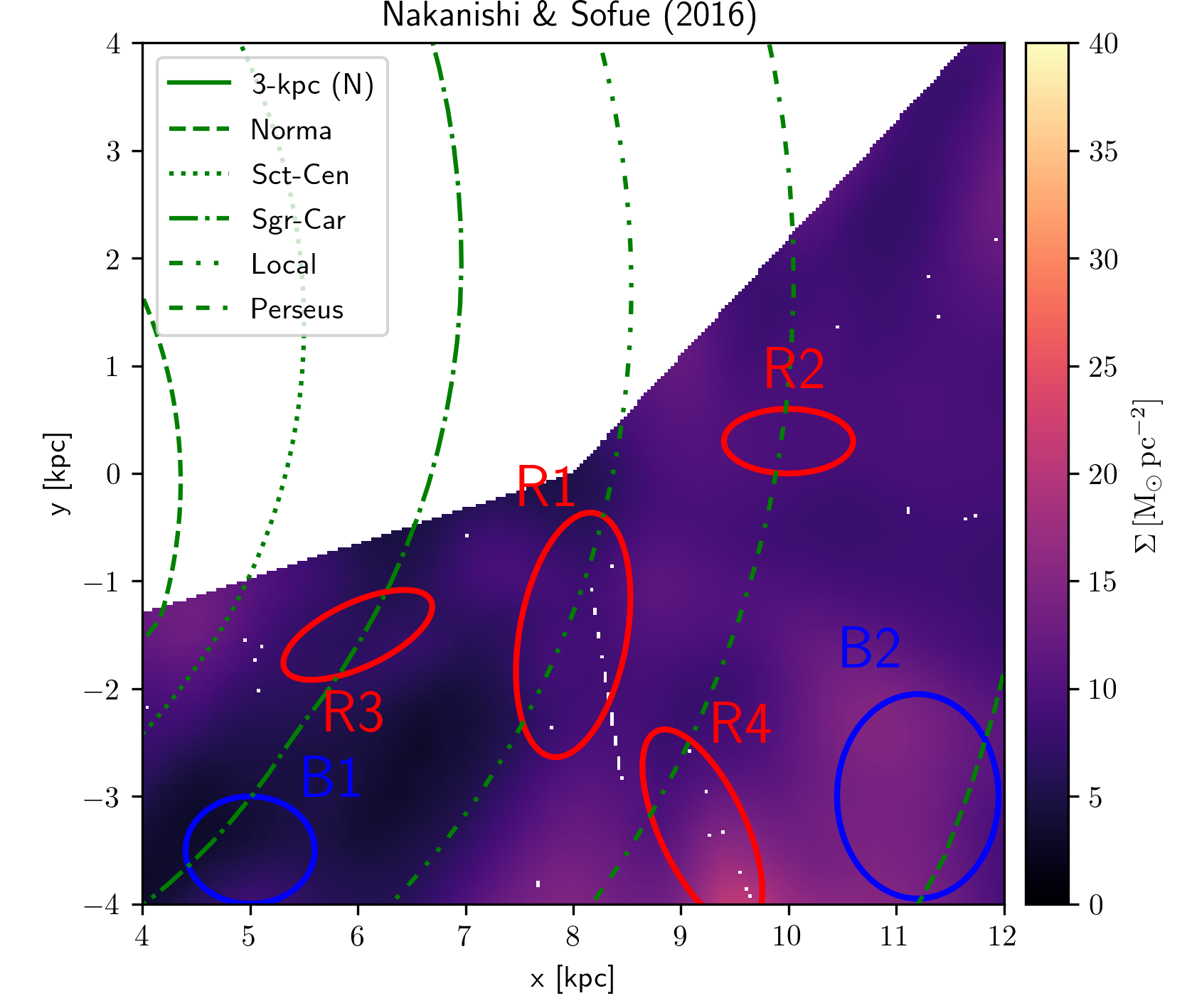}\vspace{0.5cm}
\caption{Surface mass density of hydrogen atoms converted from the reddening maps of \citet{green2019} (top left) and \cite{leike2022} (top right). Surface mass density of hydrogen atoms reconstructed from HI4PI data using the \citetalias{bissantz2003} and \citetalias{sormani2015} gas flow models (middle left and right panels respectively). Surface mass density of hydrogen atoms reconstructed by \citet{nakanishi2016} (bottom panel). The Galactic spiral structure fitted from data of molecular masers as in \citet{reid2019} is also overlaid on all the plots. The red (blue) ellipses mark some representative regions in which structures from our gas maps are seen (are not seen) in the dust maps.}
\label{fg:dust-gas}
\end{figure*}

Even though we would at this point not adopt the dust reddening data as a part of the input for our Bayesian analysis because of the issues mentioned above, it might be interesting to provide some qualitative comparisons between the gas map converted from dust reddening with our gas reconstructions. We took the three-dimensional maps of the dust reddening $E(B-V)$ from both \citet{green2019} and \cite{leike2022} and converted them into three-dimensional maps of hydrogen atoms (both atomic and molecular) using the simple relation provided in \citet{peek2013} (see also Eq. 21.6 of \citealt{draine2011}),
\begin{eqnarray}
\frac{N({\rm HI})+2N({\rm H}_2)}{E(B-V)}=7\times 10^{21} \,{\rm H}\,{\rm cm}^{-2}\,{\rm mag}^{-1}
\end{eqnarray}
The relation between the differential dust reddening and the gas density becomes 
\begin{eqnarray}
\frac{{\rm d}E(B-V)}{{\rm d}s}=7\times 10^{21} \,{\rm H}\,{\rm cm}^{-2}\,{\rm mag}^{-1} \left[n({\rm HI})+2n({\rm H}_2)\right]\n\\
\end{eqnarray}
We note that this relation has implicitly assumed a relation between the reddening and the amount of dust and, more importantly, a dust-to-gas ratio that might vary at different positions in our Galaxy \citep{tchernyshyov2017}.

The surface mass density of hydrogen atoms converted from dust reddening data is presented in the top panels of Fig.~\ref{fg:dust-gas}. We shall refer to the maps derived from \citet{green2019} and \citet{leike2022} respectively as the \citetalias{green2019} and \citetalias{leike2022} maps. The reconstruction by \citet{nakanishi2016} is also shown in Fig.~\ref{fg:dust-gas} (bottom panel) together with our reconstructions from the \citetalias{bissantz2003} and \citetalias{sormani2015} gas flow models (middle left and right panels). Note that we combined the \HI gas maps from this work with the H$_2$ gas maps from \citet{mertsch2021b} to produce the \citetalias{bissantz2003} and \citetalias{sormani2015} maps of all hydrogen atoms. We also overlay the spiral structures fitted with data of molecular masers associated with very young high-mass stars, \citep{reid2019}.

One can actually spot many structures which appear in both in the \citetalias{green2019} and in \citetalias{leike2022} maps and our reconstructions. We have marked some of these structures with red ellipses, for example, the segment of gas following the Local arm (roughly stretching between $(8,0)$ kpc and $(7,-3)$ kpc in the gas map, region R1) or the clump of gas at around $(10,0.3)$ kpc (region R2). Some other similarities exist as well, but they might be less obvious and differ slightly among these maps. For instance, the clump of gas at $(6,-1.5)$ kpc on the Sgr-Car arm (region R3) in the \citetalias{green2019} map seems to stretch along the line of sight similar to the corresponding structures in the \citetalias{bissantz2003} and \citetalias{sormani2015} maps but it has lower surface density than the corresponding structures in our reconstructions. The \citetalias{leike2022} map shows, in contrast, that the structure in the R3 region is clustered around the Sgr-Car arm and does not stretch along the line-of-sight, but it has a higher surface density that is comparable to the surface density of the corresponding structure in our reconstructions. These differences and similarities seem to hold true also for the clump of gas around $(9,-3)$ kpc on the Perseus arm (region R4). We also see that some structures are specific to one of these maps, especially on very small scales ($\lesssim 1$ kpc). Such differences might be related to the differences in the three-dimensional reconstructed dust maps and the uncertainties of the gas flow model. We believe, however, that the majority of structures in the \citet{green2019} and \citet{leike2022} maps could be identified in the gas maps for the region within roughly 3 kpc from the Solar System.

At larger heliocentric distance, the conversion might be not straightforward between dust and gas. 
Yet, there are a few clusters of gas density that are likely misplaced in the \citetalias{bissantz2003} map, possibly also in the \citetalias{sormani2015} map. The clearest such example is possibly in the region B2 at a distance of $\sim 6 \, \text{kpc}$, see Fig.~\ref{fg:dust-gas}. In the \citetalias{bissantz2003} map, there is significant gas density along longitude $134^{\circ}$. The \citetalias{sormani2015} map is based on a different rotation curve and places this emission at a much closer distance of $3$ - $4 \, \text{kpc}$. The maser distance along this direction is $\sim 2 \, \text{kpc}$~\citep{reid2019} instead.

\end{appendix}

\end{document}